\documentclass[journal]{IEEEtran}
\usepackage{dsfont}

\usepackage{color}
\usepackage{setspace}
\usepackage{clrscode}
\usepackage{amsthm}
\usepackage{pict2e}
\usepackage{amsmath}
\usepackage{amssymb}
\usepackage{amsthm}
\usepackage{graphicx}
\usepackage{stfloats}
\usepackage{soul}  
\soulregister\cite7 
\usepackage{cite}

\usepackage{algorithm}
\usepackage{algorithmic}
\usepackage{caption}
\usepackage{subcaption}
\usepackage{bbold}
\usepackage{mathabx}
\usepackage[ps2pdf,
bookmarks=false,
bookmarksnumbered=false, 
bookmarksopen=false, 
colorlinks=false]{}

\makeatletter

\newcommand{\Rmnum}[1]{\expandafter\@slowromancap\romannumeral #1@}
\makeatother

\newcommand{\x}{{\bf{x}}}
\newcommand{\y}{{\bf{y}}}
\newcommand{\Fbar}{{\overline{F}}}

\newcommand{\cL}{\mathcal{L}}
\newcommand{\Rpdf}{{\text{Ricepdf}}}

\newcommand{\sinr}{{\text{{SINR}}}}
\newcommand{\los}{{\text{{LOS}}}}
\newcommand{\nlos}{{\text{{NLOS}}}}
\newcommand{\texp}{{\text{{exp}}}}
\newcommand{\tsinr}{{\text{SINR}}}

\newcommand{\prob}{\mathbb {P}}

\newcommand{\E}{\mathbb{E}}

\newcommand{\R}{{\mathbf{R}}}

\newcommand{\mK}{\mathcal{K}}

\newtheorem{Lem}{Theorem}
\newtheorem{Lemm}{Lemma}
\newtheorem{Rem}{Remark}

\newtheorem{Corr}{Corollary}

\allowdisplaybreaks[2]
\begin{document}


%
\title{Uplink Coverage in Heterogeneous mmWave Cellular Networks with Clustered Users}
\author{\IEEEauthorblockN{Xueyuan Wang and M. Cenk Gursoy}
	\thanks{The authors are with the Department of Electrical
		Engineering and Computer Science, Syracuse University, Syracuse, NY, 13244
		(e-mail: xwang173@syr.edu, mcgursoy@syr.edu).}
	\thanks{The material in this paper was presented in part at the 2018 IEEE Vehicular Technology Conference (VTC) - Fall, Chicago, Aug. 2018.}}
\maketitle



	
\begin{abstract}
A $K$-tier heterogeneous mmWave uplink cellular network with clustered user equipments (UEs) is considered in this paper. In particular, UEs are assumed to be clustered around small-cell base stations (BSs) according to a Gaussian distribution, leading to the Thomas cluster process based modeling. Specific and practical line-of-sight (LOS) and non-line-of-sight (NLOS) models are adopted with different parameters for different tiers. The probability density functions (PDFs) and complementary cumulative distribution functions (CCDFs) of different distances from UEs to BSs are characterized. Coupled association strategy and largest long-term averaged biased received power criterion are considered, and general expressions for association probabilities are provided. Following the identification of the association probabilities, the Laplace transforms of the inter-cell interference and the intra-cluster interference are characterized.  Using tools from stochastic geometry, general expressions of the SINR coverage probability are provided. As extensions, fractional power control is incorporated into the analysis, tractable closed-form expressions are provided for special cases, and average ergodic spectral efficiency is analyzed.  Via numerical and simulation results, analytical characterizations are confirmed and the impact of key system and network parameters on the performance is identified.
\end{abstract}


\thispagestyle{empty}




\section{Introduction}
Demand for mobile data has been growing rapidly in recent years resulting in a global bandwidth shortage for wireless service providers \cite{Milli_TSP,Milli_SunR}. For future cellular networks, two key techniques for capacity improvement will be network densification and the use of higher frequencies such as in millimeter wave (mmWave) bands \cite{Uplink_HElshaer}. Indeed, with this motivation, the mmWave frequency band ranging from 30-300 GHz has recently attracted much interest for deployment in next-generaton wireless networks \cite{Milli_MRA, Milli_survey_hemadeh2017}. According to \cite{Milli_ZhouP}, the available spectrum for cellular communications in mmWave bands can be easily 200 times greater than the spectrum presently allocated for that purpose below  3 GHz. Due to this, fifth generation (5G) cellular networks will  operate in the high-bandwidth underutilized mmWave frequency spectrum \cite{Milli_survey_uwaechia}.

At the same time, certain challenges exist in realizing mmWave communications, such as severe path loss, sensitivity to blockage, directivity, and narrow beamwidth, due to the short wavelengths \cite{Milli_survey_uwaechia, Milli_survey_wang2018,Milli_survey_rappaport2017}.  For instance, the critical difficulties in using mmWave frequencies are their high path loss and very poor penetration into buildings. As a result of the blockage effect associated with mmWave, outdoor mmWave base stations (BSs) are more likely to serve outdoor user equipments (UEs) \cite{Milli_AVAlejos}. The indoor coverage in this case can be provided by other means such as indoor mmWave femtocell or Wi-Fi solutions \cite{Milli_TSP}.  On the other hand, with the use of large antenna arrays and  beamforming at the transmitter and receiver, such as in mmWave massive multiple-input-multiple-output (massive MIMO) systems, frequency-dependent path and penetration losses can be compensated \cite{Milli_ZhouP, Milli_SKutty,Milli_survey_busari2017, Milli_survey_ahmed2018}. Multiple antennas can be utilized in order to accomplish a multiplexing gain, a diversity gain, or an antenna gain, thus enhancing the bit rate, the error performance, or the signal-to-noise-pluse-interference ratio (SINR) of wireless systems \cite{Mlli_JMietzner}.

 As another trend, heterogeneous cellular wireless networks are being developed to support higher data rates to satisfy the increasing UE demand for broadband wireless services, by supporting the coexistence of denser but lower power small-cell BSs with the conventional high-power and low density macrocell BSs \cite{Uplink_ASakr,cluster_Yuang,Milli_HSDhillon,Milli_HSJo}. While macro BSs were deployed fairly uniformly to provide a ubiquitous coverage blanket, the small-cell BSs are deployed to complement capacity of the cellular networks or to patch the coverage dead zones \cite{cluster_CSahaPCP}.  A common approach in modeling the locations of BSs in each tier and the UEs is to use  independent Poisson point processes (PPPs). However, in high-density areas and hotspots, UEs are very likely to be clustered, e.g., in coffee shops, bookstores, subway/bus stations, sports/concert centers. Moreover, PPP-based models will not correctly reflect the locations of the non-uniformly distributed UEs.  In such cases, it is important to accurately capture not only the non-uniformity but also the coupling across the locations of the UEs and small-cell BSs \cite{cluster_MAf}.
 Among different point processes, Poisson cluster processes (PCPs) have been shown to lead to realistic and accurate models for characterizing the statistical nature of user-centric BS deployments and clustered UE distributions in urban areas \cite{cluster_YZhou}.
 Indeed, PCPs have been frequently employed in the literature in order to represent the correlation between the locations of UEs and BSs, which are placed at the cluster centers \cite{cluster_CSaha}. The third generation partnership project (3GPP) has also addressed the clustered configurations in which the locations of the UE and small-cell BSs  are coupled, in addition to the uniformly distributed UEs \cite{cluster_CSahaPCP_conf}.


\subsection{Related Studies}
As discussed above, we need to incorporate the relations between the locations of small-cell BSs and UEs locations. A simply way of achieving this is to use PCPs, which is also quite consistent with the 3GPP configurations \cite{cluster_CSahaPCP}. In \cite{cluster_CSahaPCP}, a unified $K$-tier heterogeneous network (HetNet) model is considered and  an arbitrary number of BS tiers and a fraction of UEs are modeled by PCPs. Coverage probability of the proposed network is analyzed and the authors have shown that the PCP weakly converges to a PPP as the cluster size of the PCPs tends to infinity. The authors in \cite{cluster_Yuang,cluster_CSaha,cluster_CSahaPCP_conf,cluster_CSaha_Unified,cluster_ullah2020performance, Cluster_Milli_WYi} have also addressed  $K$-tier HetNet models with clustered UEs. In \cite{cluster_Yuang}, the transmitting nodes are clustered following a PCP, and outage and coverage probabilities of the network are investigated. In \cite{cluster_CSaha}, the UE locations are considered as PCP distributed with the BSs at the cluster centers, and the downlink coverage probability is analyzed. Specific PCPs, namely, Thomas cluster process and Mat\'ern cluster processes are considered.  \cite{cluster_CSahaPCP_conf} modeled a fraction of UEs and several BS tiers as PCP distributed, indicating that within a macro cell the UEs can be either uniformly distributed or clustered and the small-cell BSs locations can be either uncorrelated or correlated. The authors have shown that the network performance is highly sensitive to the assumptions on the UE and small-cell BS configurations.  A two-tier HetNet consisting of macrocell and small-cell BSs is considered in \cite{cluster_MAf}, where the small-cell BSs and UEs are clustered around the geographical centers of UE hotspots. In \cite{cluster_CSaha_Unified}, the BSs' locations in each tier of the HetNet are modeled as a PPP or PCP, and the locations of the UEs are modeled using either  a uniform distribution or a PCP.  \cite{cluster_ullah2020performance} considered a HetNet, which contained the union of uniformly distributed UEs and  UEs distributed according to  a Thomas cluster process. 
A clustered mmWave network in which non-orthogonal multiple access (NOMA) techniques are employed, is introduced in \cite{Cluster_Milli_WYi}, where the NOMA UEs are modeled as PCP distributed and each cluster contains a BS located at the center. 
The authors in \cite{cluster_chen2018coverage} considered a downlink user-centric wireless network in a comprehensive fading environment,  where the locations of the UEs are modeled by a Thomas cluster process.

Several recent studies have also addressed the uplink analysis in mmWave networks.
In \cite{Uplink_OOnireti}, a framework to evaluate the SINR coverage in the uplink of mmWave cellular networks with fractional power control (FPC) is presented. Conventional path loss based FPC and distance based FPC are considered. The locations of line-of-sight (LOS) UEs and non-LOS (NLOS) UEs are modeled as two independent non-homogeneous PPPs which are independent of the locations of BSs.
A hybrid network with traditional sub-6 GHz macrocells coexisting with  mmWave small-cells is addressed in \cite{Uplink_MShi} and \cite{Uplink_HElshaer}.  The authors in \cite{Uplink_MShi} have analyzed the decoupled downlink and uplink association strategies. In \cite{Uplink_HElshaer}, different decoupled uplink and downlink cell association strategies are investigated based on two different criteria, namely maximum biased received power and maximum achievable rates. 
Similarly as in mmWave studies, a path loss model incorporating both LOS and NLOS transmissions in uplink of dense small-cell networks is considered in \cite{Uplink_TDing}. Additionally, the UE density is assumed to be higher than the BS density, while they are still spatially distributed according to independent PPPs.
Moreover, the energy efficiency maximization problem was investigated for uplink mmWave systems with non-orthogonal multiple access (NOMA) in   \cite{uplink_MZeng_energy, uplink_XYu_joint}. 

Moreover, uplink analysis in PCP distributed networks has recently been conducted in several studies. For instance,
device-to-device (D2D) enabled mmWave cellular networks with clustered UEs are analyzed in \cite{Milli_WYi} and \cite{Cluster_D2D_Esma} , transceivers are modeled according to a PCP in \cite{Milli_WYi}, and both PPP-distributed cellular UEs and PCP-distributed D2D UEs are considered in \cite{Cluster_D2D_Esma}. Coverage probability and area spectral efficiency of the networks are analyzed in both papers.
The authors in \cite{cluster_uplink_HTabassum} have provided a framework to analyze single-tier multi-cell uplink NOMA systems where the UE locations are modeled following a Mat\'ern cluster process. Rate coverage probability of NOMA users and the mean rate coverage probability of all users in a cell are characterized. \cite{Unlink_arif2020decoupled} considered a decoupled downlink and uplink access scenario in a HetNet, where the distribution of UEs are modeled as a Matern cluster process. Closed-from expressions for system coverage probability, spectral efficiency and energy efficiency were obtained. 
We note that these prior studies have not considered uplink analysis in mmWave networks.


\subsection{Main Contributions}
Motivated by the fact that mmWave communications and clustered UE deployments have been attracting growing attention as significant components in next-generation wireless networks, we in this paper study $K$-tier heterogeneous uplink mmWave cellular networks with clustered UE deployments.
The key contributions of this paper can be summarized as follows:

\begin{itemize}
	\item We present a mathematical framework for evaluating the performance of multi-tier heterogeneous uplink mmWave cellular networks. Compared to downlink analysis, analyzing the uplink performance presents challenges due to the difficulty in accurately characterizing the interference from UEs. We address these challenges in this paper and conduct an uplink analysis.

\item Unlike the papers considering uniformly distributed UEs, in this paper we concentrate on the non-uniformly distributed UEs, and we employ PCPs to accurately model the locations of the clustered UEs. In particular, Thomas cluster process is adopted, with which the UEs are clustered around the small-cell BSs according to a Gaussian distribution. We also address the coexistence of small-cell BSs and marcocell BSs.
	\item  The distinguishing properties of mmWave network, i.e., directional antenna gains and different path loss models for LOS and NLOS links, are incorporated into the analysis.  We derive the Laplace transforms of the inter-cell interference and the interference from the cluster member UE (the intra-cluster interference).
    Note that since UE transmissions inflict interference in uplink, characterization of the interference requires an analysis different from that in the downlink case \cite{Milli_XWang_journal}. Additionally, since we focus on the clustered UEs in this paper, determining the Laplace transform of the interference is also different compared to scenarios in which UEs are modeled as PPP.

\item Following the characterization of the interference, distance distributions and association probabilities, we derive general expressions for the SINR coverage probability.
	\item We extend the analysis to incorporate Nakagami fading and also consider several special cases, e.g., the noise-limited case,  interference-limited case, and one-tier model. Furthermore, we provide extensions to include fractional power control in the analysis and reformulate the Laplace transforms of the interference in this case. The extensions indicate the generality of the main analysis, which can be easily modified to adapt to new network models. Average ergodic spectral efficiency of the network is also analyzed.
\end{itemize}

 The following insights are gained from numerical and simulation results:  1) when the standard deviation of the UE distribution approaches infinity, the performance of our PCP-based model converges to that of the PPP-based model; 2) the type of small-scale fading does not have significant influence on SINR coverage; 3) interference has considerable impact on network coverage, especially when fractional power control is employed by the UEs; 4) the SINR coverage performance strongly depends on the standard deviation of the UE distribution, the LOS probability function exponent, and the density of the small-cell BSs; 5) the SINR coverage probability can be improved by increasing the biasing factor and transmit power of small-cell BSs.

The remainder of the paper is structured as follows: In Section II, the system model is introduced. In Section III, the characterizations of the distances between UEs and BSs from each tier are provided. In Section IV, the association probabilities of each tier are derived.  In Section V, the Laplace transforms of the inter-cell interference are characterized,  and  general expressions for the SINR coverage probability in each tier are derived. In Section VI, extensions and special cases are addressed. In Section VII, numerical results are presented, and association probabilities and coverage probabilities with respect to different system parameters are analyzed. Finally, in Section VIII, conclusions are drawn. Proofs are relegated to the Appendix.

\section{System Model} \label{SM}
In this section, we introduce the system model of the considered network. BS and UE spatial distributions, and the channel models are described in detail.
\subsection{Base Station Distribution Modeling}
We consider a $K$-tier heterogeneous uplink mmWave cellular network. $\mK =\{1,2,...,K\} $ is used to denote the set of tier indices. BSs in the $i^{th}$ tier are distributed according to the independent homogeneous PPP $\Phi_i$.
BSs in different tiers differ in spatial density $\lambda_i$, transmit power $P_i$ and biasing factor $B_i$ (which is used to describe the association priority), and they are assumed to transmit in mmWave frequency bands. Out of $K$ tiers, it is assumed that there are $K_u$ tiers of small-cell BSs, around which UEs are clustered, and we denote the set as $\mK_u$.

\subsection{User Equipment Distribution Modeling}
Unlike existing works which mostly consider UEs distributed uniformly according to some independent homogeneous point process, we consider a network scenario where the UEs are clustered around the small-cell BSs. In each cluster, small-cell BS and the UEs are regarded as the cluster center and cluster members, respectively. Assume that the cluster center is a BS from the $i^{th}$ tier, then the cluster members are regarded as the $i^{th}$ tier UEs, and then we can use $\mK_u$ denote the set of tier indices for UEs. The union of the cluster members form a PCP, denoted as $\Phi^{i}_u$. In this paper $\Phi^{i}_u$ is modeled as a Thomas cluster process (TCP) where cluster members are spatially distributed around the cluster center BS according to a Gaussian distribution with standard deviation $\sigma_i$. Initially, in the main analysis of this paper, we do not consider power control, and the UEs are assumed to transmit at the same power level of $P_u$. Subsequently, we extend the analysis to incorporate fractional power control in Section VI-B.

\begin{figure}
	\begin{minipage}{0.48\textwidth}
		\centering
		\label{fig:subfig:b}
		\includegraphics[width=1 \textwidth]{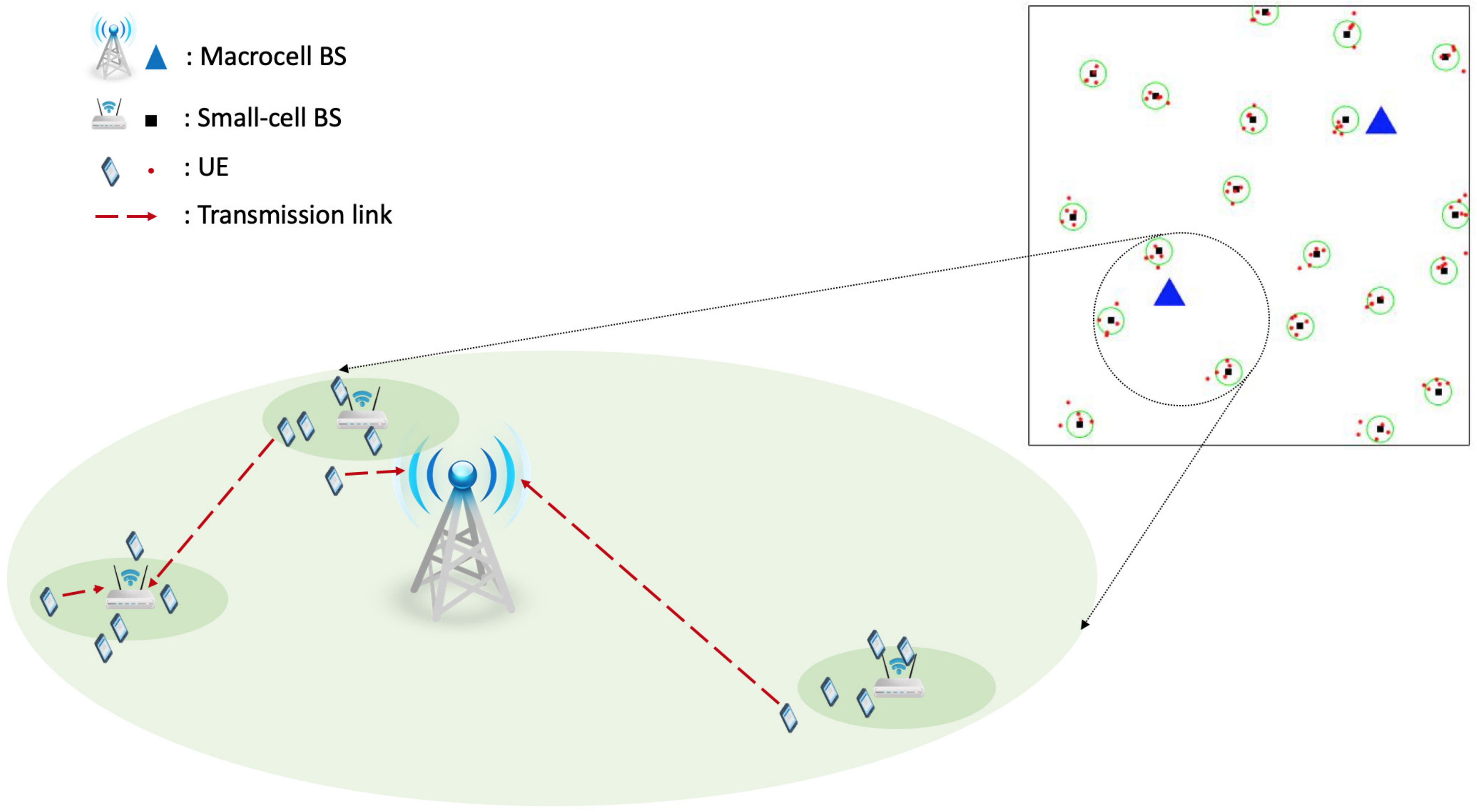}
		\subcaption{\scriptsize Users in Thomas cluster process. }
	\end{minipage}
	\caption{\small Two-tier heterogeneous network model, where macrocells (blue triangles) and small-cells ( black squares) are distributed as independent PPPs.  
	 UEs (red dots) are distributed around small-cells according to a Gaussian distribution. The average number of UEs per cluster is 5. \normalsize}
	\label{PPP_TCP}
\end{figure}

We use $0^{th}$ tier to stand for the cluster center to the UEs, or the cluster members to the BSs.
A two-tier heterogeneous network models are depicted in Fig. \ref{PPP_TCP}, providing illustration of  the TCP model considered in this paper.

\subsection{Directional Beamforming}
In mmWave networks, BSs and UEs are equipped with directional antenna arrays to compensate for the high path loss. We assume that the antenna arrays at the BSs and UEs perform directional beamforming. For simplicity, we adopt a widely used sectored antenna pattern model as in \cite{Uplink_MShi,Milli_Bai,Milli_JAndrews}, where $M_*, m_*, \theta_* $ (where the subscript $*\in\{b,u\}$ indicates a BS or a UE) are used to denote the main lobe gain, side lobe gain and the beamwidth of the main lobe, respectively.  It is assumed that the typical BS and its associated UE can adjust their antenna steering orientation after estimating the angle of transmission and angle of reception, and achieve the largest antenna gain $G_0= M_b M_u$. Beam direction is assumed to be independently and uniformly distributed between [0, 2$\pi$) (if no beam alignment is considered). Hence, the antenna gain between the typical BS and the interfering UEs can be expressed as
 \begin{align} \label{antennaG}
   G= \left\{
            \begin{array}{ll}
               M_b M_u  & \text{with prob. } p_{M_bM_u} = \left(\frac{\theta_b}{2\pi} \right) \left(\frac{\theta_u}{2\pi} \right) \\
               M_b m_u  & \text{with prob. } p_{M_bm_u} =  \frac{\theta_b}{2\pi} \left(1-\frac{\theta_u}{2\pi}\right)\\
               m_b M_u  & \text{with prob. } p_{m_bM_u} =  \left(1-\frac{\theta_b}{2\pi}\right) \frac{\theta_u}{2\pi} \\
               m_b m_u  & \text{with prob. } p_{m_bm_u} = \left(1-\frac{\theta_b}{2\pi}\right)\left(1-\frac{\theta_u}{2\pi}\right)
            \end{array}
          \right.
 \end{align}
 where $p_G$ denotes the probability of having antenna gain $G \in \{M_b M_u, M_b m_u,m_b M_u,m_b m_u \}$.

\subsection{Channel Model}
Blockage models are adopted in mmWave networks to characterize the high near-field path loss and poor penetration through solid materials. A mmWave link between a BS and a UE is assumed to be either LOS or NLOS, where LOS indicates that no blockage exists and UEs are visible to the BS, while NLOS implies that blockage exists between the UE and the connecting BS. The path loss function can be expressed as
\begin{align}
L^s_i(r)=
\left\{
\begin{array}{ll}
\kappa_L r^{\alpha_L}  & \text{with prob. } p^L_{\alpha i} \\
\kappa_N r^{\alpha_N}  & \text{with prob. } p^N_{\alpha i}=1-p^L_{\alpha i}
\end{array}
\right.
\end{align}  
where $\alpha_L$ and $\alpha_N$ are the LOS and NLOS path loss exponents, respectively; and $\kappa_L$ and $\kappa_N$ are the intercepts in the LOS and NLOS path loss. In a network, to characterize the LOS probability function, field measurements \cite{Milli_MRA} or stochastic blockage models \cite{Channel_MFranceschetti} can be used. When the blockages are modeled as a rectangle Boolean scheme in \cite{Milli_TBai2},  the probability of LOS link between the BS and UE can be formulated as
\begin{align}
	p^L_{\alpha i} (r) = e^{-\epsilon r}
\end{align}
where $\epsilon$ is a constant that depends on the geometry and density of the building blockage process. All transmission links are assumed to experience independent Rayleigh fading with unit mean, i.e., channel fading power is exponentially distributed,  $h \sim \exp(1)$. However, our analysis can be extended by incorporating Nakagami fading for the main transmission link, and this extension is addressed in Section VI-A.

\begin{table}[htbp]
	\caption{Table of Notations}
	\centering
	\begin{tabular}{|l|p{2.5in}|}
		\hline
		\footnotesize \textbf{Notations} &  \footnotesize  \textbf{Description}  \\ \hline
		
		\scriptsize$K, \mK, \mK_u$ &  \scriptsize Total number of tiers of BSs, the set of all tiers of BSs,  the set of all tiers of UEs.  \\ \hline	
		\scriptsize$\Phi_i, \lambda_i $&  \scriptsize PPP of BSs in $i^{th}$ tier, and the density. \\ \hline	
		\scriptsize$\Phi^i_u, \sigma_i$&\scriptsize  PCP of UEs in $i^{th}$ tier, and the standard deviation. \scriptsize  \\ \hline		
		\scriptsize$P_i,  B_i, P_u$&  \scriptsize Transmit power and the biasing factor of BSs in $i^{th}$ tier, and transmit power of UEs. \\ \hline	
		\scriptsize$G, p_G$  &\scriptsize Effective antenna gain and the corresponding probability. \\ \hline	
		\scriptsize$M,m, \theta$  &\scriptsize  Main lobe directivily gain, side lobe gain, and the beamwidth of the main lobe. \\ \hline				
		\scriptsize$\alpha_s, \kappa_s$ &\scriptsize  Path loss exponent and intercept  of an $s\in \{\los,\nlos\}$ link.          \\ \hline
		\scriptsize$L^s_i, p^s_{\alpha i}$ &\scriptsize Path loss and the probability  of an $s$ link in the $i^{th}$ tier.         \\ \hline
		\scriptsize$ \epsilon$ &\scriptsize A constant that depends on the geometry and density of the building blockage process.        \\ \hline
		\scriptsize$ h, \sigma^2_{n}$ &\scriptsize Small-scale fading.   The variance of the noise component.     \\ \hline
		\scriptsize$r_{i0} $ &\scriptsize The distance from a UE to its cluster center BS.       \\ \hline
		\scriptsize$r^s_{j} $ &\scriptsize The distance from a UE to the nearest $j^{th}$ tier LOS/NLOS BS.       \\ \hline
		\scriptsize$r_{jk} $ &\scriptsize The distance from a $k^{th}$ tier UE to a $j^{th}$ tier BS.       \\ \hline
		\scriptsize$f_r, \Fbar_r $ &\scriptsize The PDF and CCDF of $r$.       \\ \hline		
		\scriptsize $S_{ij,s}$   &   \scriptsize An event  that an $i^{th}$ tier UE is associated with a $j^{th}$ tier BS with an $s\in \{\los,\nlos \}$ transmission.  \\ \hline		
		\scriptsize $A_{ij,s}$   &   \scriptsize The probability of event $S_{ij,s}$. \\ \hline
		\scriptsize $\sinr_{ij,s}$   &   \scriptsize The SINR experienced at a reference $j^{th}$ tier BS, if it serves an $i^{th}$ tier UE with $s$ transmission. \\ \hline
		\scriptsize $I_{jk}$   &   \scriptsize The interference, experienced at a reference $j^{th}$ tier BS, from the $k^{th}$ tier UEs. \\ \hline
		\scriptsize $\cL_I(\mu)$   &   \scriptsize Laplace transform of $I$ evaluated at $\mu$. \\ \hline
		\scriptsize $T$   &   \scriptsize Target SINR threshold.  \\ \hline	
		\scriptsize $P^c_{C_{ij,s}}$   &   \scriptsize The SINR coverage probability given $S_{ij,s}$. \\ \hline
		\scriptsize $P_C$   &   \scriptsize The uplink SINR coverage probability of the network.   \\ \hline	
		\scriptsize $\mathcal{R}$   &   \scriptsize Average ergodic spectral efficiency.   \\ \hline			
	\end{tabular}
\end{table}\normalsize

\section{Distance characterization}
\begin{figure}
	\centering
	\includegraphics[width=0.35 \textwidth] {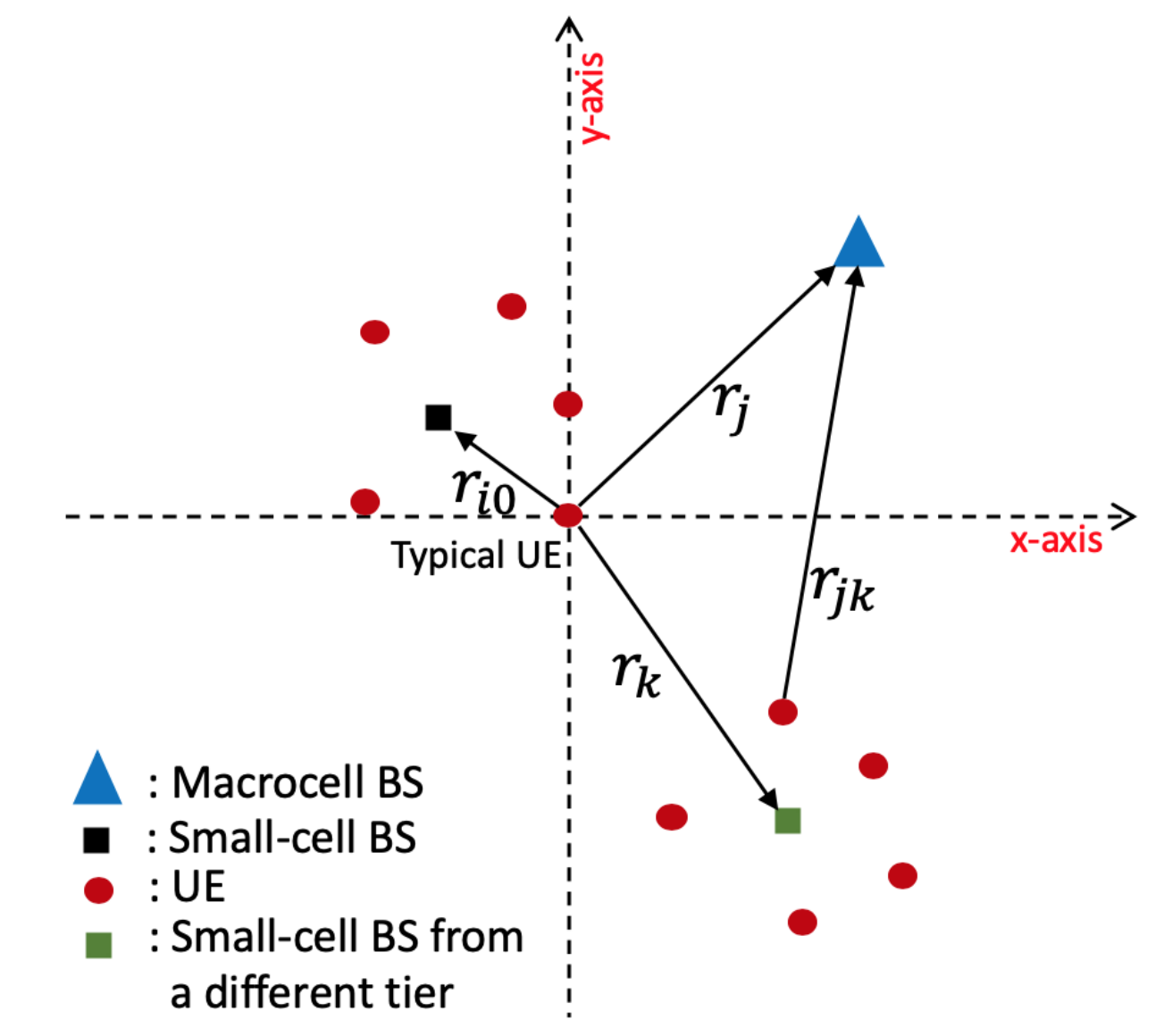}
	\caption{\small An illustration of different distances from the BSs to UEs. \normalsize}
	\label{distanceModel}
\end{figure}

In this section, we statistically describe the distances between the BSs and UEs.  A representative scenario displaying different distances is shown in Fig. \ref{distanceModel}. Assume that the typical UE is from the $i^{th}$ tier and is located at the origin. 
 The probability density function (PDF) and complementary cumulative distribution function (CCDF) of these distances are discussed below, and will be used in the analysis of association probability and the system performance matrices.

\subsection{Distance from the typical UE to its cluster center BS $r_{i0}$ ($i\in \mK_u$) }
Since UEs are scattered around their cluster centers according to a Gaussian distribution, the PDF of $r_{i0}$ can be expressed as \cite{cluster_MAf_Model}
\begin{align} \label{ri0PDF}
f_{r_{i0}}(x) = \frac{x}{\sigma_i^2} \, \texp \left(\frac{-x^2}{2 \sigma_i^2} \right)  \qquad (x\geq 0)
\end{align}
and the CCDF is
\begin{align} \label{ri0CCDF}
\Fbar_{r_{i0}}(x) = \exp \left(\frac{-x^2}{2 \sigma_i^2}\right)          \qquad (x\geq 0)
\end{align}
where $\sigma_i$ is the standard deviation of UE distribution $\Phi_u^i$. Note that this distance is Rayleigh distributed.

Given the link is in $s\in \{\los,\nlos\}$ channel, the PDF and CCDF of $r_{i0}$ can be expressed as \cite{UAV_XWang}
\begin{align}
&f_{r^s_{i0}}(x) = p^s_{\alpha i}(x) f_{r_{i0}}(x) /D^s_{i0}   \\
&	\Fbar_{r^s_{i0}}(x) = \int_{x}^{\infty} p^s_{\alpha i}(x) f_{r^s_{i0}}(t)  dt /D^s_{i0}
\end{align}
where $D^s_{i0} = \E_{r_{i0}}[p^s_{\alpha i}(r_{i0})]$ is the probability that the link between the UEs and their cluster cluster center small-cell BSs is in $s$ channel.


\subsection{Distance from the typical UE (the origin) to the nearest $j^{th}$ tier LOS/NLOS BS $r^s_{j}$ ($j \in \mK$) }
Given that the typical UE can observe at least one LOS/NLOS BS in the  $j^{th}$ tier, the PDF and CCDF of the distance can be expressed as \cite{Milli_Bai}
\begin{align}
&\label{rjPDF}f_{r^s_{j}}(x)=2 \pi \lambda_j x p^s_{\alpha j}(x) e^{-2 \pi \lambda_j \int^{x}_0 tp^s_{\alpha j}(t)dt  }  /D^s_{j} \\
&\label{rjCCDF}\Fbar_{r^s_{j}}(x)=e^{-2 \pi \lambda_j \int^{x}_0 tp^s_{\alpha j}(t)dt  }  /D^s_{j}
\end{align}
where $D^s_{j}=1-e^{-2 \pi \lambda_j \int^{\infty}_0 x p^s_{\alpha j}(x) dx}$ is the probability that the typical UE has at least one $s\in \{\los, \nlos\}$ $j^{th}$ BS around.


\subsection{Distance from the $k^{th}$ tier UE to the $j^{th}$ tier BSs $r_{jk}$ ($j\in \mK$, $k\in\mK_u$)}
For a UE from the $k^{th}$ tier, the distance from the cluster center BS  to the $j^{th}$ tier BSs is notated by $v_{jk}$
Since the cluster members are distributed according to a Gaussian distribution with variance $\sigma_k^2$, the conditional distribution of $r_{jk}$ given $v_{jk}$ is a Rice distribution, and the PDF is given by \cite{cluster_MAf_Model}
\begin{align} \label{rice}
&f_{r_{jk}}(r| v_{jk}) = \Rpdf(r,v_{jk}; \sigma_k^2)   \qquad  \quad (r\geq 0)
\end{align}
where  Ricepdf$(a,b;\sigma^2)=\frac{a}{\sigma^2} \exp(-\frac{a^2+b^2}{2\sigma^2}) I_0(\frac{ab}{\sigma^2})$ and $I_0(\cdot)$ is the modified Bessel function of the first kind with order zero. This distance formulation is used in the analysis of interference.

%

\section{Association Probability} \label{AP}
In this section, we first describe the downlink-uplink coupled association strategy, then provide expressions for the association probability and discussions on the conditional PDF of the distance from the typical UE to the reference BSs. Since the typical UE can be served by its cluster center BS, other small-cell BSs, and the macrocell BSs, we have three types of reference BSs in the network. Association probability describes the probability that the typical UE is served by a reference BS. In Section V, we investigate the coverage probabilities at each reference BS, and together with the association probability we eventually investigate the coverage performance of the entire network.

\subsection{Association Criterion}
 Coupled association strategy is adopted in this paper, which constrains the serving BS to be the same in both uplink and downlink.  More specifically, all BSs send pilot signals to the UEs in downlink transmission, and the UEs choose the BS providing the largest long-term averaged biased received power \cite{Uplink_MShi} \cite{Milli_XWang_journal} to connect in both downlink and uplink transmissions.  
 	Here, long-term averaging indicates that we average over the small-scale fading. Biasing factor describes the association priority of each UEs to different tiers of BSs, and is denoted by $B_j$ for the $j^{th}$ tier BSs. Larger $B_j$ indicates that UEs have higher preference to be associated with $j^{th}$ tier BSs.
  This largest long-term averaged biased received power at a UE can be formulated as
\begin{align}\label{P1}
  P=\mathop{\max}_{j\in (\mK \cup 0),n\in\Phi_j}{P_{j,n} B_{j,n} G_0 L_{j,n}^{-1}}
\end{align}
where ${P_{j,n}, B_{j,n},L_{j,n}}$ are the transmit power, biasing factor, and the path loss of the $n^{th}$ BS in the $j^{th}$ tier, respectively. $G_0$ is the effective antenna gain. Since in the $j^{th}$ tier, the transmit power and the biasing factor are the same, the BS in this tier with the minimum path loss to the UE provides the maximum received power $P$. Now, we can rewrite
\begin{align}\label{P2}
  P=\mathop{\max}_{j\in(\mK \cup 0)}{P_{j} B_{j} G_0 L_{j,min}^{-1}}
\end{align}
where $L_{j,min}= \kappa r_{min}^{\alpha}$ is the minimum path loss in the $j^{th}$ tier.

\subsection{Association Probability}
Based on the coupled association strategy, we can define the association probability (denoted by $A_{ij,s}$) as the probability  that an $i^{th}$ tier UE is served by a $j^{th}$ tier BS in $s$ channel, indicating that this BS provides the largest long-term averaged biased received power to the typical UE in downlink transmission.
 We define the event $S_{ij,s}$ = \{the typical UE from the $i^{th}$ tier is served by a BS from the $j^{th}$ tier with a $s \in \{\los, \nlos\}$ link\}. Again, we note that the typical UE can be served by  its cluster center BS and in this case we set $j=0$.


  \begin{figure*}[bp]
	\hrulefill	
	\begin{align}
	A_{ij,s} =
	\begin{cases}
	D^s_{i0} \E_{r^s_{i0}} \left[ \prod_{\substack{k=1}  }^K \prod_b D^b_{k}  \Fbar_{r^b_{k}} [(C^{bs}_{k0}{r^s_{i0}}^{\alpha_s})^{\frac{1}{\alpha_b}}] \right], &j=0, \\
	\E_{r^s_j}\left[ D^s_{j} D^{s'}_{j} \Fbar_{r^{s'}_{j} }\left( \frac{\kappa_{s'}}{\kappa_{s}} {r^s_{j}}^{\frac{\alpha_s}{\alpha_{s'}}}\right) \Big[ \sum_{a} D^a_{i0} \Fbar_{r^s_{i0}}( (C^{as}_{0j} {r^s_{j}}^{\alpha_s})^{\frac{1}{\alpha_a}})  \Big]
	\left[\prod_{\substack{k=1 \\ k\neq i}  }^K \prod_{b} D^b_{k} \Fbar_{r^b_{k}} ((C^{bs}_{kj} {r^s_{j}}^{\alpha_s})^{\frac{1}{\alpha_b}}) \right] \right], &j\in\mK,
	\end{cases}
	\label{AP}  	
	\end{align}
   \end{figure*}
  The probability that the typical UE from the $i^{th}$ tier is associated with a BS from the $j^{th}$ tier in a LOS/NLOS transmission can be obtained by utilizing \cite[Lemma 3]{UAV_XWang}, and the expressions are given in (\ref{AP}) below at the bottom of the next page for two different cases.
	In (\ref{AP}), $s,s',a,b\in \{\los, \nlos\}$,  $C^{sb}_{kj}=\frac{P_k B_k \kappa_s}{P_j B_j \kappa_b}$, $D^b_{k}=1-e^{-2 \pi \lambda_k \int^{\infty}_0 x p^b_{\alpha k}(x) dx}$ is the probability that a UE has at least one $b\in \{\los, \nlos\}$ $k^{th}$ tier BS around, $D^s_{i0} = \E_{r_{i0}}[p^s_{\alpha i}(r_{i0})]$  is the probability that the link from the typical UE to its cluster center BS is in $s \in \{\los, \nlos \}$ condition, and $\Fbar_r(x)$ is the CCDF of $r$ given in (\ref{ri0CCDF}) and (\ref{rjCCDF}) in Section III.



\subsection{Conditional PDF of the distance from the typical UE to the associated BS $r_{ij,s}$ given $S_{ij,s}$}
The conditional CDF of $r_{ij,s}$ given $S_{ij,s}$ can be expressed as
\begin{align}
 \hat{F}_{r_{ij,s}}(x) &= \prob(r_{ij,s}<x |S_{ij,s}) \notag \\
 &=  \frac{\prob(r_{ij,s}<x ,S_{ij,s})}{\prob(S_{ij,s})} = \frac{\prob(r_{ij,s}<x ,S_{ij,s})}{A_{ij,s}}.
\end{align}
Following a similar approach as in the analysis of the association probability, we obtain the conditional PDF as
\begin{align}
&\hat{f}_{r_{ij,s}} (x)= \frac{d\hat{F}_{r_{ij,s}}(x)}{dx} =\notag \\
&\begin{cases}
\frac{f_{r^s_{i0}} (x) }{A_{i0,s}} D^s_{i0} \prod_{\substack{k=1}  }^K \prod_b D^b_{k}  \Fbar_{r^b_{k}} [(C^{bs}_{k0}{x}^{\alpha_s})^{\frac{1}{\alpha_b}}], \quad j=0, \\
\frac{f_{r^s_j} (x)}{A_{ij,s}} D^s_{j} D^{s'}_{j} \Fbar_{r^{s'}_{j} }\left( \frac{\kappa_{s'}}{\kappa_{s}} {x}^{\frac{\alpha_s}{\alpha_{s'}}}\right) \Big[ \sum_{a} D^a_{i0} \Fbar_{r^s_{i0}}( (C^{as}_{0j} {x}^{\alpha_s})^{\frac{1}{\alpha_a}})  \Big] \\
\hspace{0.6in} \left[\prod_{\substack{k=1 \\ k\neq i}  }^K \prod_{b} D^b_{k} \Fbar_{r^b_{k}} ((C^{bs}_{kj} {x}^{\alpha_s})^{\frac{1}{\alpha_b}}) \right] , j\in\mK.
\end{cases}
\end{align}

\section{Coverage Analysis} \label{CP}
In this section, coverage probabilities are analyzed. First by utilizing tools from stochastic geometry, we provide general expressions for coverage probabilities. Subsequently, the Laplace transforms of the inter-cell interference and intra-cluster terms are characterized for Thomas cluster processes.

\subsection{Signal-to-interference-plus-noise Ratio (SINR)}
Due to orthogonal resource allocation among users in a cell, there is only one UE from each outside cell causing interference to a reference BS.
We note that if the reference BS is the cluster center BS to the typical UE, this BS only experiences inter-cell interference; otherwise the reference BS experiences both inter-cell and the interference from a cluster member UE (intra-cluster interference).
We model the interfering UEs as $\Phi'_{uk}$.

Therefore, the interference can be expressed as follows:
\begin{align}
       &\text{Inter-cell interference: } I_{jk}=\sum\limits_{n\in\Phi'_{uk}} P_u G_{k,n} h_{k,n}  \kappa^{-1} r_{k,n}^{-\alpha} \\
       &\text{Intra-cluster interference: } I_{j0} = P_u G h \kappa^{-1} r_{j0} ^{-\alpha}
\end{align}
where $P_u$ is the transmit power of the UEs,  and $G_{k,n}$ and $ h_{k,n}$ are the effective antenna gain and the small-scale fading gain of the $n^{th}$ interfering UE in $\Phi'_{uk}$, respectively. We note that if the reference BS is the cluster center BS, $I_{j0}=0$.

The SINR experienced at the reference BS can be expressed as
\begin{align}
        \sinr_{ij,s}= \frac{P_u G_0 h_j \kappa_s r_{}^{-\alpha_s}}{\sigma^2_{n} + I_{j0} + \sum\limits_{k\in\mK_u} I_{jk}}
\end{align}
where $s\in \{\los, \nlos \}$, $G_0= M_b M_u$ and  $\sigma^2_{n}$ is the variance of the additive white Gaussian noise component.

\subsection{SINR Coverage Probability}
The association probability is the probability that the typical UE is served by a reference BS.
In this subsection, we analyze the  SINR coverage probability $ P^c_{C_{ij,s}}$ which is defined as the probability that the experienced SINR at the reference BS is above a certain threshold $T > 0$ given $S_{ij,s}$, i.e., $ P^c_{C_{ij,s}}= \prob(\sinr_{ij,s}>T|S_{ij,s})$. The coverage probability of the entire network can be formulated as
    \begin{align}
    P_C &=\sum_{j=0}^K P_{C_{ij}} =  \sum_{j=0}^K \sum_{s\in \{L,N \}} \prob(\sinr_{ij,s}>T|S_{ij,s}) \prob(S_{ij,s}) \notag \\
    & = \sum_{j=0}^K \sum_{s\in \{L,N \}}  P^c_{C_{ij,s}} A_{ij,s}
    \label{CP_whole}
    \end{align}
    where $A_{ij,s}$ is the association probability given in (\ref{AP}). The following result characterizes the coverage probabilities.

\begin{Lem} \label{theo:coverage}
Given the event $S_{ij,s}$
, the SINR coverage probability is formulated as
\begin{align}
P^c_{C_{ij,s}} &=\prob(\tsinr_{ij,s}>T_j | S_{ij,s}) \notag \\
&=\E_{r_{ij,s}} \left[ e^{-\mu^s_{ij} \sigma^2_{n }} \cL_{I_{j0}} (\mu^s_{ij}) \prod\limits_{k=1}^{K_u} \cL_{I_{jk}} (\mu^s_{ij})  \right],
\label{CP_conditional}
\end{align}
where $s \in \{\los, \nlos \}$, $\mu_{ij}^s= \frac{T_j \kappa_s r_{ij,s}^{\alpha_s}}{P_u G_0 } $, $I_{jk}$ is the interference from the UEs in the $k^{th}$ tier to the reference BS in the $j^{th}$ tier, $I_{i0}$ is the interference to the reference BS from its cluster members. $\cL_I(\mu)=\E\{\exp(-\mu I)\}$ is the Laplace transform of I evaluated at $\mu$, and $\cL_{I_{i0}}(\mu)=1$ when $I_{i0}=0$.

Proof: See Appendix \ref{Proof_CP}.
\end{Lem}

\subsection{Laplace Transforms of the Interference Terms}

In this subsection, we provide characterizations for the Laplace transforms of the interference terms.

\begin{Lem}
If the UEs are distributed around small cell BSs according to Gaussian distributions, the Laplace transform of the interference is given by the following:\\
1) If the interference arises from the cluster member, we have
\begin{align}
    &\cL_{I_{j0}}(\mu_{ij}^s)=\sum_G \sum_{a\in \{L,N\}} p_G D^a_{j0} \cL_{I_{j0}^{Ga}}(\mu_{ij}^s)  \label{Lap1}
\end{align}
where
\begin{align}
    &\cL_{I_{j0}^{Ga}}(\mu_{ij}^s)=
      \int_0^\infty \frac{1}{1+ \mu_{ij}^s P_u G \kappa^{-1}_a r_{j0}^{-\alpha_a}} f_{R^s_{j0}}(r_{j0}) dr_{j0}.
      \label{Lap3}
\end{align}
2) If the interference arises from the UEs in the $k^{th} (k\in \mK_u)$ tier, we have
 \begin{align}
    & \cL_{I_{jk}}(\mu_{ij}^s) = \prod_G \prod_{a \in \{L, N \} }  \cL_{I_{jk}^{Ga}}(\mu_{ij}^s)  \label{Lap2}
\end{align}
where
\begin{align}
    &\cL_{I_{jk}^{Ga}}(\mu_{ij}^s)= \notag \\
    &e^{ -2 \pi \int_0^{\infty} \lambda_k p_G  p^a_{\alpha k}(r_{jk,n})\left( \frac{1}{1+( \mu_{ij}^s P_u G \kappa^{-1}_a r_{jk,n}^{-\alpha_a})^{-1}} \right) r_{jk,n} dr_{jk,n} }. \label{Lap4}
\end{align}
Above, $G \in \{M_bM_u,M_bm_u,m_bM_u,m_bm_u \} $, $s, a \in \{\los, \nlos \}$,  is the antenna gain, $p_G$ is probability of having different antenna gains.

Proof: See Appendix \ref{Proof_Laplace}.
\end{Lem}

\section{Extensions and Special cases}
In this section, we extend our analysis to incorporate Nakagami fading and also fractional power control. We also address average ergodic spectral efficiency. Finally, we consider special cases for which we obtain simplified expressions for the association and coverage probabilities.

\subsection{Extension to Nakagami Fading }
It is well known that Rayleigh fading is a special case of Nakagami fading and is obtained by setting the Nakagami parameter $N_s=1$. Hence, Nakagami is a more general fading distribution that provides a better fit to several practical scenarios and experimental results. In this subsection, we extend the coverage analysis to Nakagami fading.

\begin{Corr}
	When small-scale Nakagami fading with Nakagami parameter $N_s$ (where $s \in \{\text{LOS, NLOS}\}$) is considered for the main link, the SINR coverage probability of each tier is given by the following:
\begin{align}
&P^c_{C_{ij,s}}= \notag \\
&	\E_{r_{ij,s}}  \left[ \sum_{n=1}^{N_s} (-1)^{n+1} \Big(\substack{N_s \\ \\n}\Big) e^{- \mu_{ij}^s \sigma_{n}^2 } \cL_{I_{j0}}(\mu_{ij}^s)   \prod\limits_{k=1}^{K_u}  \cL_{I_{jk}}(\mu_{ij}^s) \right], \label{CP2p}
	\end{align}
	where when $j=0$, $I_{j0}=0$ and $ \cL_{I_{j0}}(\mu_{ij}^s) =1$.
	In the above coverage probability expressions, the Laplace transforms of the interference terms follow (\ref {Lap1})-(\ref{Lap4}) with the same expressions.
\end{Corr}

\emph{Proof:} The proof follows along the same lines as in the proof of Theorem \ref{theo:coverage}. The only change is that the moment generating function (MGF) of the now gamma-distributed fading gain  $h$ is applied in the computation of the expectation $\E_h [\cdot]$.

\subsection{Extension to UE fractional power control}

Two types of fractional power control can be considered for UEs:  1) path loss based fractional power control is performed by compensating the path loss of the UE irrespective of
whether its path to the serving BS is LOS or NLOS; 2) distance based fractional power control is performed by inverting with the LOS path loss exponent \cite{Uplink_OOnireti}. The transmit power of UEs can be expressed as follows:
\begin{align}
P'_u =
\begin{cases}
P_u t^{\tau \alpha_L }, & \text{if distance based,} \\
P_u L^{\tau},  & \text{if path loss based,} \\
\end{cases}
\label{power_control}
\end{align}
where $\tau \in [0,1]$ is the power control factor, $t$ is the distance from the UE to its serving BS, $\alpha_L$ is the LOS path loss exponent, and $L$ is the path loss of the link. When $\tau =0$, no power control is performed. We consider the path loss based fractional power control as in LTE \cite{Uplink_powercontrol_ASimonsson}.

With fractional power control, the general expression for SINR coverage probability essentially remains the same as in (\ref{CP_whole}) and (\ref{CP_conditional}), with only $P_u$ replaced by $P'_u$. However, the Laplace transform of the interference should be modeled differently.
Let us use $I_{jk}^{Ga}$ to denote the interference from the $k^{th}$ tier UEs to a $j^{th}$ tier reference BS. Since the transmit power of UEs depend on the path loss from its serving BS (the cluster center BS, small-cell BSs, or macro BSs), we need to distinguish different cases considering the transmit power of interfering UEs. Therefore, we have
\begin{align}
I_{jk}^{Ga} = \sum_{\substack{m=0 \\ m\neq j}  }^{K} (I_{jkmL}^{Ga} + I_{jkmN}^{Ga})
\end{align}
where interference $I_{jkmL}^{Ga}$ occurs when the $k^{th}$ tier interfering UEs are served by an $m^{th}$ tier BS over a LOS link. Next we characterize the Laplace transforms of the interference terms.

\begin{Lemm}
1) If the interference arises from the cluster member, we have
	\begin{align}
	&\cL_{I_{j0}}(\mu_{ij}^s)=\sum_G \sum_{a\in \{L,N\}} \sum_{m=1}^{K} \sum_{b \in \{L, N \} } p_G D^a_{j0} A_{km,b} \cL_{I_{j0mb}^{Ga}}(\mu_{ij}^s)
	\end{align}
	where
	\begin{align}
	\cL_{I_{j0}^{Ga}}(\mu_{ij}^s)
	=\int_0^\infty \int_0^\infty \frac{\hat{f}_{r_{km,b} } (t )f_{R^s_{j0}}(r_{j0})}{1+ \mu_{ij}^s P_u (\kappa_b t^{\alpha_b})^{\tau} G \kappa^{-1}_a r_{j0}^{-\alpha_a}}  dt dr_{j0}.
	\end{align}
2) If the interference arises from the UEs in the $k^{th} (k\in \mK_u)$ tier, we have
	\begin{align}
	& \cL_{I_{jk}}(\mu_{ij}^s) = \prod_G \prod_{a \in \{L, N \} }\prod_{\substack{m=0 \\ m\neq j}  }^{K} \prod_{b \in \{L, N \} }  \cL_{I_{jkmb}^{Ga}}(\mu_{ij}^s)
	\end{align}
	where
	\begin{align}
	&\cL_{I_{jkmb}^{Ga}}(\mu_{ij}^s)= \notag \\
	& e^{ -2 \pi A_{km,b}  \lambda_k p_G   \int_0^{\infty} \int_{0}^{\infty}    \frac{p^a_{\alpha k}(r_{jk,n}) \hat{f}_{r_{km,b} (t) r_{jk,n} }}{1+( \mu_{ij}^s (\kappa_b t^{\alpha_b})^{\tau} G \kappa^{-1}_a r_{jk,n}^{-\alpha_a})^{-1}} dt  dr_{jk,n} },
	\end{align}
	where $\mu_{ij}^s= \frac{T_j (\kappa_s r_{ij,s}^{\alpha_s})^{1-\tau}}{ P_u G_0 } $, $t$ is the distance from the interfering UE to its own associated BS, and $r_{jk,n}$ is the distance from the interfering UE to the reference BS.
	
	Proof: See Appendix \ref{Proof_PowerControl}.
\end{Lemm}

%

\subsection{Average ergodic spectral efficiency}
According to \cite{Uplink_OOnireti} \cite{Milli_TBai2}, given the SINR coverage probability $P_C(T_j)$,
the average ergodic spectral efficiency of the uplink of a mmWave
cellular network can be expressed as
\begin{align}
	\mathcal{R} &= \frac{1}{\ln 2} \int_{0}^{\infty} \frac{P_C(T)}{1+T} dT \notag \\
	& =  \frac{1}{\ln 2} \int_{0}^{\infty} \frac{\sum_{j=0}^K \sum_{s\in \{L,N \}} A_{ij,s} P^c_{C_{ij,s}}(T)}{1+T} dT  \notag \\
	&= \sum_{j=0}^K \sum_{s\in \{L,N \}} \frac{A_{ij,s}}{\ln 2} \int_{0}^{\infty} \frac{P^c_{C_{ij,s}}(T)}{1+T} dT \notag \\
	& = \sum_{j=0}^K \sum_{s\in \{L,N \}} A_{ij,s} \mathcal{R}^c_{ij,s}.
\end{align}
The average achievable rate can be computed as $W\mathcal{R}$, where $W$ is the bandwidth assigned to a UE.

%
%

\subsection{Special cases}
To simplify the general expressions provided in the previous sections, we now address several special cases.
\subsubsection{Noise-limited scenario}
When the mmWave network is not densely deployed and the impact of interference is negligible (due to blockages experienced in mmWave bands), the network can be assumed to be noise-limited. In this case, the coverage probability of each tier can be simplified to
\begin{align}
P^c_{C_{ij,s}} =\E_{r_{ij,s}}  \left[ e^{- \mu_{ij}^s \sigma_{n}^2 }    \right].
\end{align}

Next we consider a further simplified case, specifically a two-tier heterogeneous network, including one tier of small-cell BSs with UEs clustered around, and one tier of macrocell BSs; all links are LOS and the path loss exponent is $\alpha=2$; and only noise is considered. Then, we can obtain the association probability and SNR coverage probability as in the following lemma.
\begin{Rem}
	In the above described simplified two-tier network model with LOS links, path loss exponent $\alpha = 2$, and no interference, we have the association and coverage probabilities given by the following:\\
	1) Association probability:
	\begin{align}
	&\label{SC_A0} A_{10}= \frac{1}{2 \sigma^2 C_0}= \frac{P_1 B_1}{P_1 B_1 + 2\pi \sigma^2 \lambda_1 P_1 B_1 +2\pi \sigma^2 \lambda_2 P_2 B_2}  \\
	&\label{SC_A1} 	A_{11}= \frac{\pi \lambda_1}{ C_0}= \frac{2\pi \sigma^2 \lambda_1 P_1 B_1}{P_1B_1 + 2\pi \sigma^2 \lambda_1 P_1 B_1 +2\pi \sigma^2 \lambda_2 P_2 B_2}  \\
	&\label{SC_A2} 	A_{12}= \frac{\pi \lambda_2}{ C_2}= \frac{2\pi \sigma^2 \lambda_2 P_2 B_2}{P_1B_1 + 2\pi \sigma^2 \lambda_1 P_1 B_1 +2\pi \sigma^2 \lambda_2 P_2 B_2}
	\end{align}
	2) SNR coverage probability:
	\begin{align}
	&\label{SC_P0}  P_{10} = \frac{1}{2 \sigma^2 (C_0 + \frac{T\sigma_n^2}{P_u G_0}) A_{10}} \\
	&\label{SC_P1} P_{11} = \frac{\pi \lambda_1 }{\left( C_0 + \frac{T\sigma_n^2}{P_u G_0} \right) A_{11}} \\
	&\label{SC_P2} P_{12} = \frac{\pi \lambda_2 }{\left(C_2 + \frac{T\sigma_n^2}{P_u G_0} \right) A_{12}}
	\end{align}
	where $\lambda_1$ and $\lambda_2$ are the densities of small-cell BSs and macrocell BSs, respectively, $\sigma^2$ is the variance of the UE distribution, $C_0= \frac{1}{2\sigma^2}+\pi \lambda_1 + \pi \lambda_2 \frac{P_2 B_2}{P_1 B_1}$, and $C_2= \frac{P_1 B_1}{2\sigma^2 P_2 B_2}+ \pi \lambda_1 \frac{P_1 B_1}{P_2 B_2} +\pi \lambda_2$.
	Proof: See Appendix \ref{Proof_NoiseLimited}.
\end{Rem}
 Results in (\ref{SC_A0}) - (\ref{SC_P2}) provide closed-form expressions for the association probabilities and SNR coverage probabilities as functions of the key system/network parameters, e.g., the standard deviation of the UE distribution $\sigma$, the transmit powers of the UEs and BSs, the densities of the BSs in each tier. From the expressions, we can directly identify the impact of these parameters on the network performance. For instance,  we note that SNR coverage probability is a monotonically increasing function of $P_u$ and $G_0$, and a monotonically decreasing function of the SNR threshold $T$. Additionally, we see that when the UEs are more and more sparsely distributed (i.e., $\sigma \rightarrow \infty$), $A_{10}$ and $P_{10}$ approach 0.

\subsubsection{Interference-limited scenario}
In cases in which the interference power satisfies $I \gg \sigma^2_n$ , the thermal noise can be ignored and the coverage probability of each tier can be expressed as
\begin{align}
&\label{SC_interferencelim}P^c_{C_{ij,s}} =\E_{r_{ij,s}}  \left[  \cL_{I_{j0}}(\mu_{ij}^s)   \prod\limits_{k=1}^{K_u}  \cL_{I_{jk}}(\mu_{ij}^s) \right].
\end{align}
In this interference-limited case, since the noise term $ e^{-\mu^s_{ij} \sigma^2_{n }}$ is removed, expression (\ref{SC_interferencelim}) is relatively easier to compute than the general case given in (\ref{CP_conditional}).

\subsubsection{Cluster center association scenario}
In the scenarios that  the clusters are sparsely distributed, e.g. in rural areas, it is reasonable to require the UEs to be associated with their cluster center BSs. In such case, the coverage probability can be expressed as
\begin{align}
&\label{SC_CC} P_{C_{i,s}}= \int\limits_{0}^{\infty} e^{-\mu^s_{i} \sigma^2_{n }} \prod\limits_{k=1}^{K_u} \cL_{I_{ik}}(\mu_{i0}^s) f_{r^s_{i0}} (r^s_{i0}) dr^s_{i0}.
\end{align}
Additionally, in this case there is no need to find the association probabilities and the conditional PDFs of the distances from UEs to the associated BSs, since UEs are connected with their cluster centers BSs. The use of $f_{r^s_{i0}} (r^s_{i0})$ makes (\ref{SC_CC}) much easier to compute than (\ref{CP_conditional}).

\section{Numerical results} \label{Num}
In this section we present numerical and simulation results to confirm our analytical characterizations and further study the performance levels in the considered mmWave network model.

In the numerical evaluations and simulations, unless specified otherwise, we consider a two-tier  heterogeneous network model  in the uplink scenario. More specifically, BSs in the $1^{st}$ tier are assumed to operate with relatively smaller transmit power but have larger density, and are regarded as small-cell BSs, while BSs in the $2^{nd}$ tier are assumed to have larger transmit power but smaller density, and are regarded as macrocell BSs. Besides, UEs are considered to cluster around small-cell BSs. Some notations are defined as follows: $A_{10}$, $P_{C_{10}}$ are the association probability (AP) that a UE is served by its cluster center BS and the coverage probability (CP) of this link, respectively; $A_{11}$, $P_{C_{11}}$ are the AP that a UE is served by a small-cell BS from other clusters and the CP of this link, respectively; and $A_{12}$, $P_{C_{12}}$ are the AP that a a UE is served by a macrocell BS  the CP of this link, respectively. Parameter values are listed in Table I below.

\begin{table}[htbp]
\small
\caption{Table of Parameter Values}
\centering
\begin{tabular}{|l|l|}
\hline
\textbf{Parameters} & \textbf{Values}  \\   \hline
\small $P_0, P_1, P_2$  & \small 30dBm, 30dBm,  46dBm \cite{Uplink_MShi} \\ \hline
\small $P_u$  &\small 23dBm \cite{Uplink_MShi}  \\ \hline
\small $B_0, B_1, B_2$  &\small  $1, 1,  1$ \\ \hline
\small$\epsilon$ &\small $\sqrt{2}/200$ \cite{Milli_Bai} \\ \hline
\small $ \alpha_d^{j,L}$, $\alpha_d^{j,N}$ $\forall j$, $\forall d$  &\small $2, 4$ \cite{Milli_Bai, Milli_Esma}\\  \hline
\small $\lambda_1, \lambda_2, \lambda_u$  &\small $10^{-4}, 10^{-5}, 5\times10^{-4}$ /m$^2$\\ \hline
\small $M, m, \theta$  &\small  10dB, -10dB,  $ \pi/6$ \cite{Milli_Bai, Milli_Esma} \\ \hline
\small Career frequency $F_c$  &\small 28 GHz \cite{Milli_Bai, Milli_Esma} \\  \hline
\small $\kappa_L=\kappa_N$   &\small $(F_c / 4 \pi)^2$ \cite{Milli_Esma} \\  \hline
\small Bandwidth $W$   &\small 100 MHz \cite{Milli_Bai} \\  \hline
\small $\sigma^2_{n,j}$ $\forall j$  &\small -174dBm/Hz + 10 log$_{10}(W)$ + 10dB  \cite{Milli_Esma}\\ \hline
\end{tabular}
\end{table}

\subsection{Impact of the cluster size}
\begin{figure}
	\centering
	\begin{minipage}{0.4\textwidth}
		\centering
		\includegraphics[width=1\textwidth]{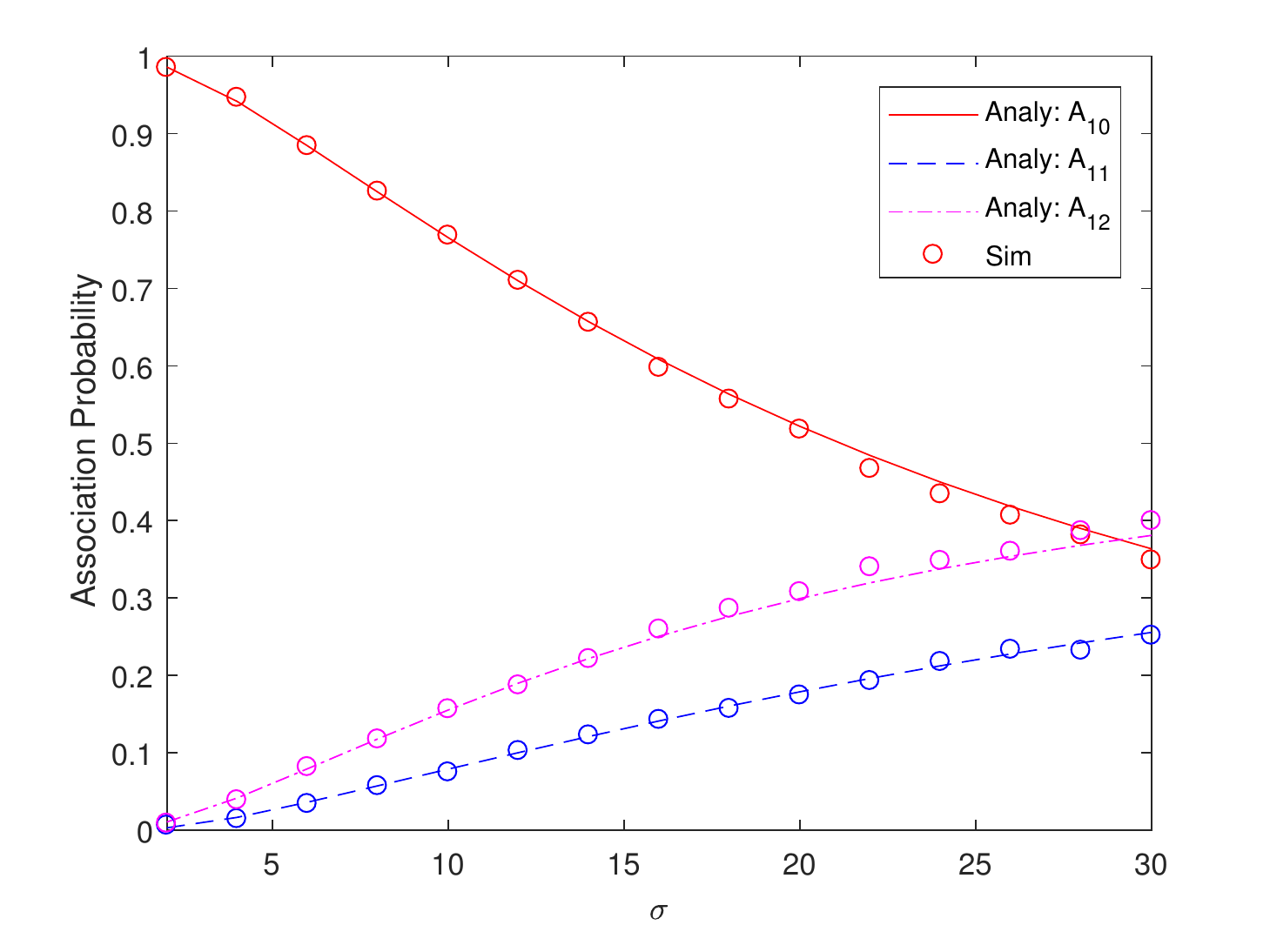} \\
		\subcaption{\scriptsize Association Probability. }
	\end{minipage}
	\begin{minipage}{0.4\textwidth}
		\centering
		\includegraphics[width=1\textwidth]{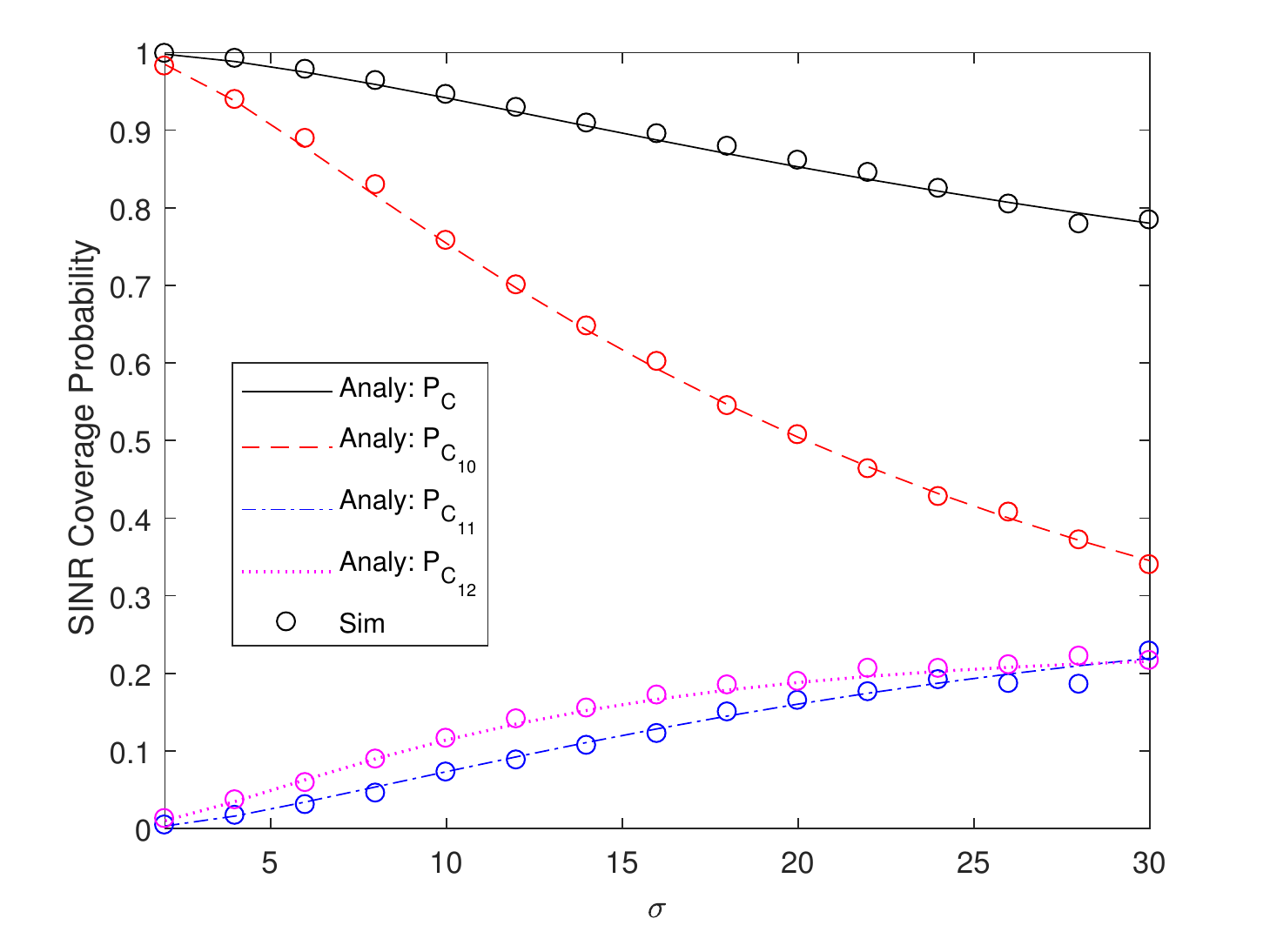}
		\subcaption{\scriptsize  SINR coverage probability.}
	\end{minipage}
	\caption{\small APs and SINR CPs as a function of the standard derivation of UE distribution when $T=10$dB. \normalsize}
	\label{AP_size}
\end{figure}

In this section, we investigate the impact of the cluster size (i.e. the standard deviation $\sigma$ of the UE distribution) on the system performance. In Fig. \ref{AP_size}, we observe that the simulation results match with the analytical results when $\sigma\leq 30$, and hence we verify our analysis. Then we can say our analysis can be applied to practical PCP-based uplink mmWave networks, providing reliable and useful insights.
\subsubsection{Association Probability (AP)}
Fig. \ref{AP_size}(a) shows the APs of each tier as a function of $\sigma$. When $\sigma$ becomes larger, UEs are getting spread further away from their own cluster center and moving closer to other BSs, which can be either small-cell or macrocell BSs. As a result, small-cell BSs become less likely to serve their cluster members and more likely to serve UEs from other clusters. Macrocell BSs become more likely to serve a UE as well. Therefore, $A_{10}$ is expected to decrease, and $A_{11}$ and $A_{12}$ to increase with the growing $\sigma$, and this is observed in Fig. \ref{AP_size}(a).

On the other hand, initially when $\sigma$ is small and therefore UEs are more tightly clustered around their cluster center, $A_{10}$ is larger than both $A_{11}$ and $A_{12}$. Also, since the macrocell BSs have much larger transmit power than small-cell BSs, $A_{12}$ is larger than $A_{11}$.

\subsubsection{Coverage Probability (CP)}
Fig. \ref{AP_size}(b) plots the SINR CP as a function of $\sigma$. Again, increasing $\sigma$ implies that the UEs are more widely distributed.
Thus, the path loss from a small-cell BS to its cluster member increases on average, while the path loss to other UEs as well as the path loss from a macrocell BS to a UE decrease. As a result, we observe in Fig. \ref{AP_size}(b)  decreasing $P_{C_{10}}$ and increasing $P_{C_{11}}$ and $P_{C_{12}}$. And since the decrease in $P_{C_{10}}$ is not compensated, the overall SINR CP diminishes as well.

\subsubsection{A three-tier heterogeneous network}
\begin{figure}
	\centering
	\begin{minipage}{0.4\textwidth}
		\centering
		\includegraphics[width=1\textwidth]{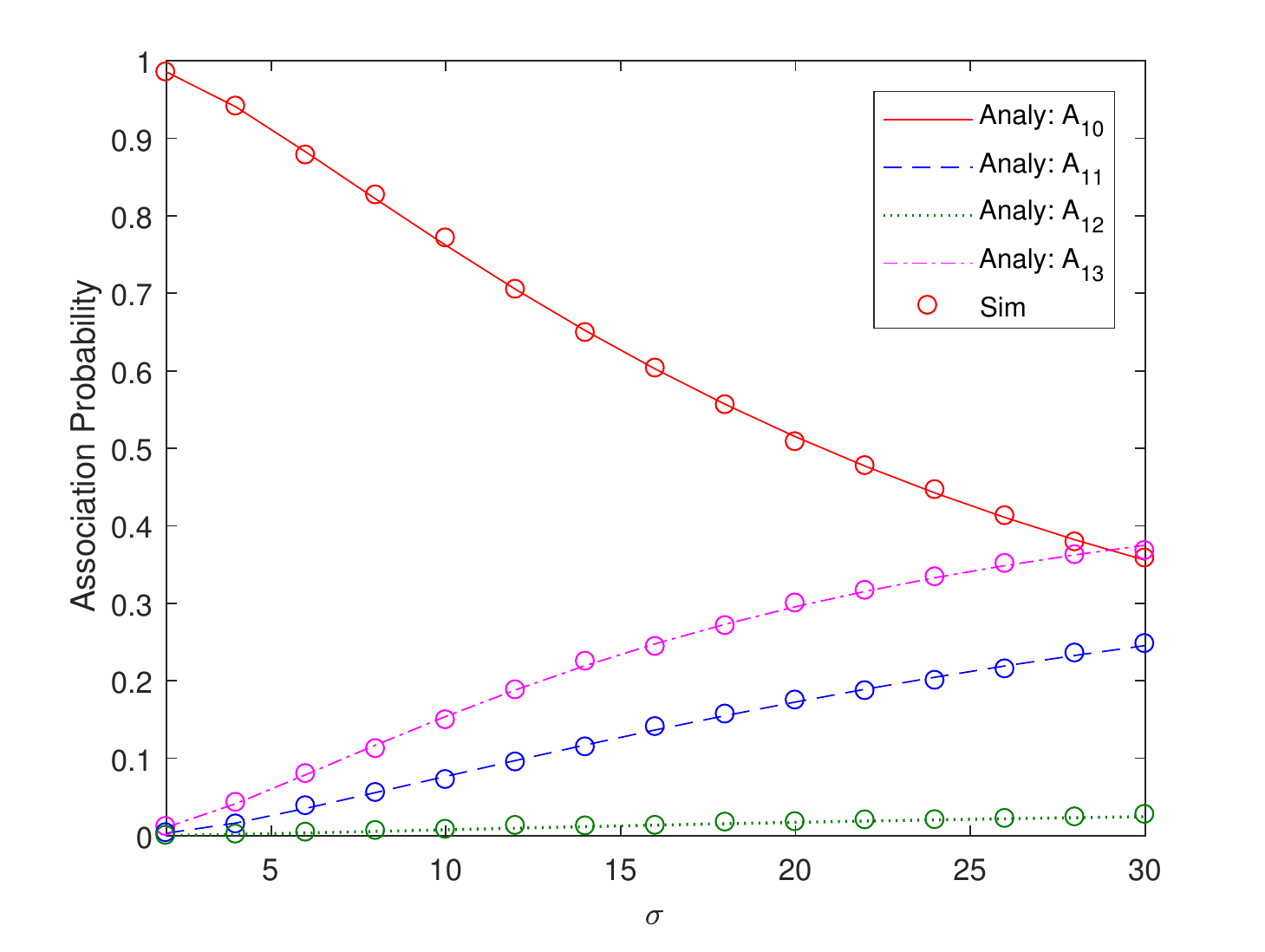} \\
		\subcaption{\scriptsize Association Probability. }
	\end{minipage}
	\begin{minipage}{0.4\textwidth}
		\centering
		\includegraphics[width=1\textwidth]{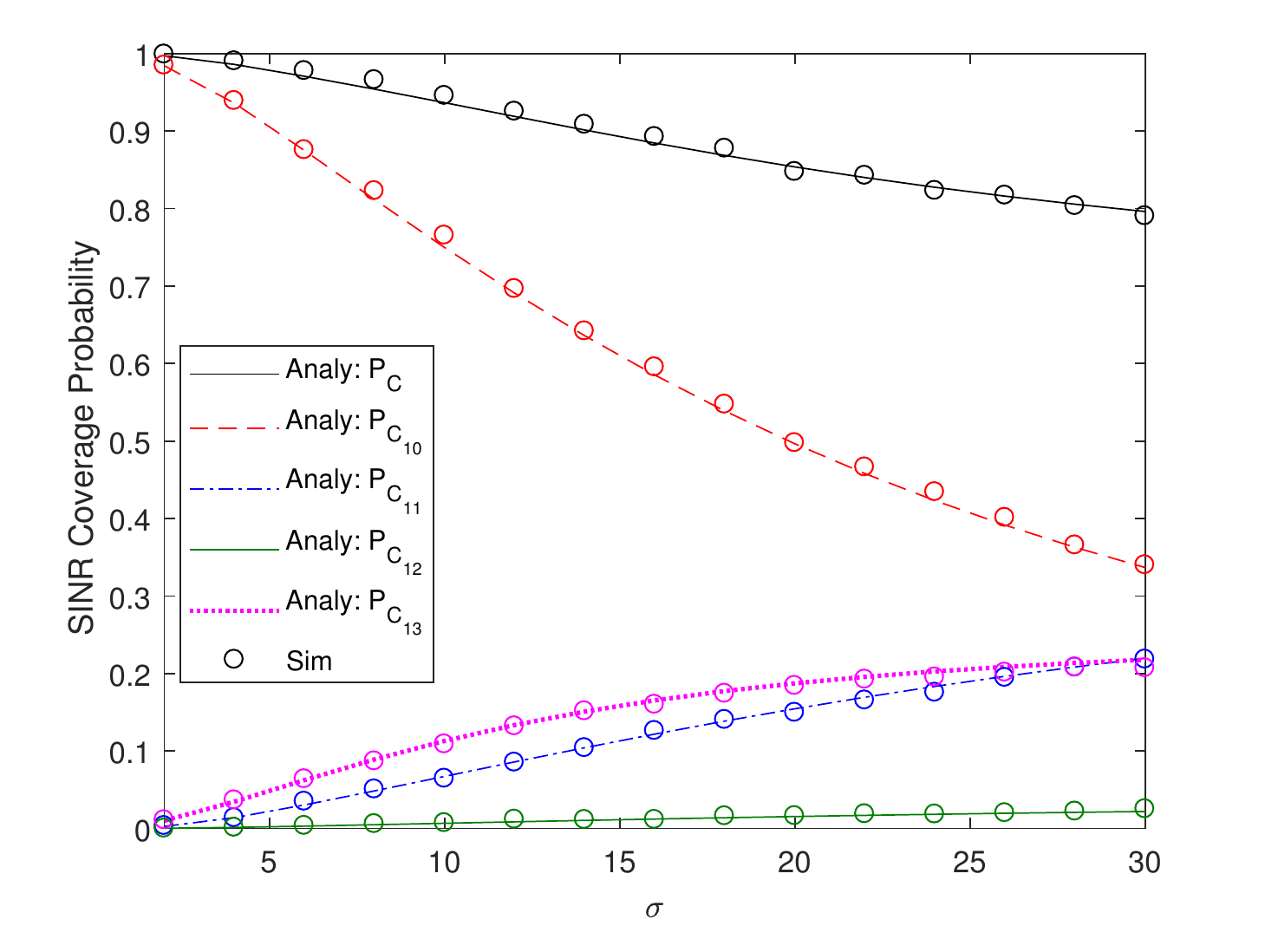}
		\subcaption{\scriptsize  SINR coverage probability.}
	\end{minipage}
	\caption{\small APs and SINR CPs as a function of $\sigma$ when $T=10$dB in a three-tier heterogeneous network. \normalsize}
	\label{Figure_3tier}
\end{figure}

Fig. \ref{Figure_3tier} plots the AP and SINR CP as a function of $\sigma$ in a three-tier heterogeneous network. In this network, we add one more tier of small-cell BSs, referred to  as the $2^{nd}$ tier with density $\lambda_2 = 10^{-5}$ /m$^2$ and UE transmission power  $P_2=$30 dBm. The macrocell BSs are now referred to as the $3^{rd}$ tier. Again, increasing $\sigma$ implies that the UEs are more widely distributed. In Fig. \ref{Figure_3tier}, we observe similar performances as in Fig. \ref{AP_size}. Since the density and transmit power of the $2^{nd}$ tier small-cell BSs are not very large, the presence of this tier does not influence the performance significantly.

\subsubsection{Comparison with PPP}
\begin{figure}
	\centering
	\includegraphics[width=0.4 \textwidth]{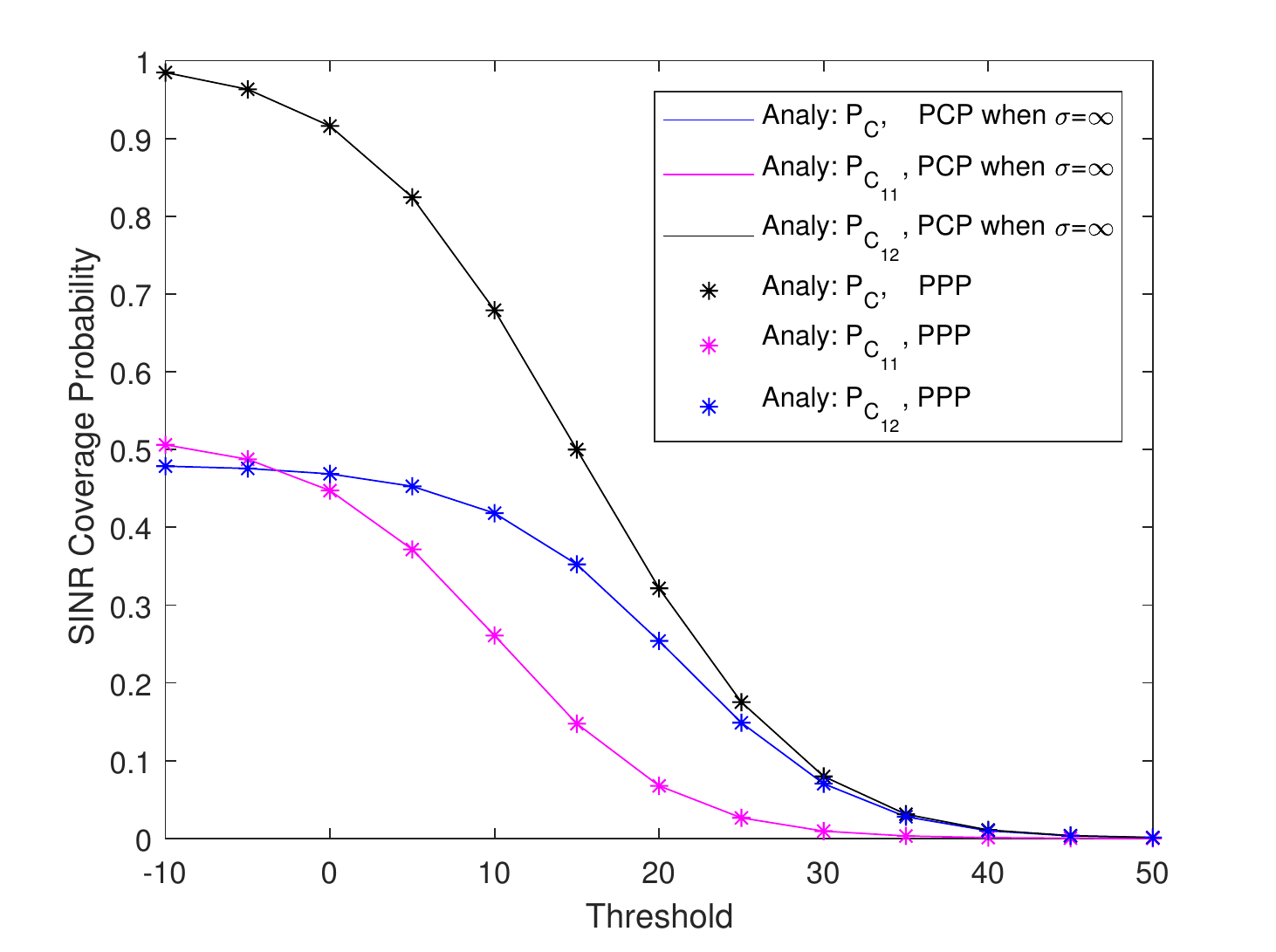}
		\caption{\small  SINR CPs as a function of threshold  $T$. \normalsize}
	\label{Figure_PCP_PPP}
\end{figure}
Fig. \ref{Figure_PCP_PPP} shows the SINR CP of the considered network when $\sigma \rightarrow \infty$ ($P_{C_{10}}=0$ is not shown in the figure) and SINR CP of the same network when the UEs are uniformly distributed according to a PPP. This figure demonstrates that as  $\sigma \rightarrow \infty$ the performance of the considered network  converges to the performance in the PPP-based model, indicating that our analysis can be specialized to determine the performance of the PPP-based model (where the density of UEs is $\phi^{i'}_u = \phi_i$) by setting $\sigma = \infty$.

\subsection{Impact of interference and the small-scale fading }
\begin{figure}
	\centering
	\begin{minipage}{0.4\textwidth}
		\centering
		\includegraphics[width=1\textwidth]{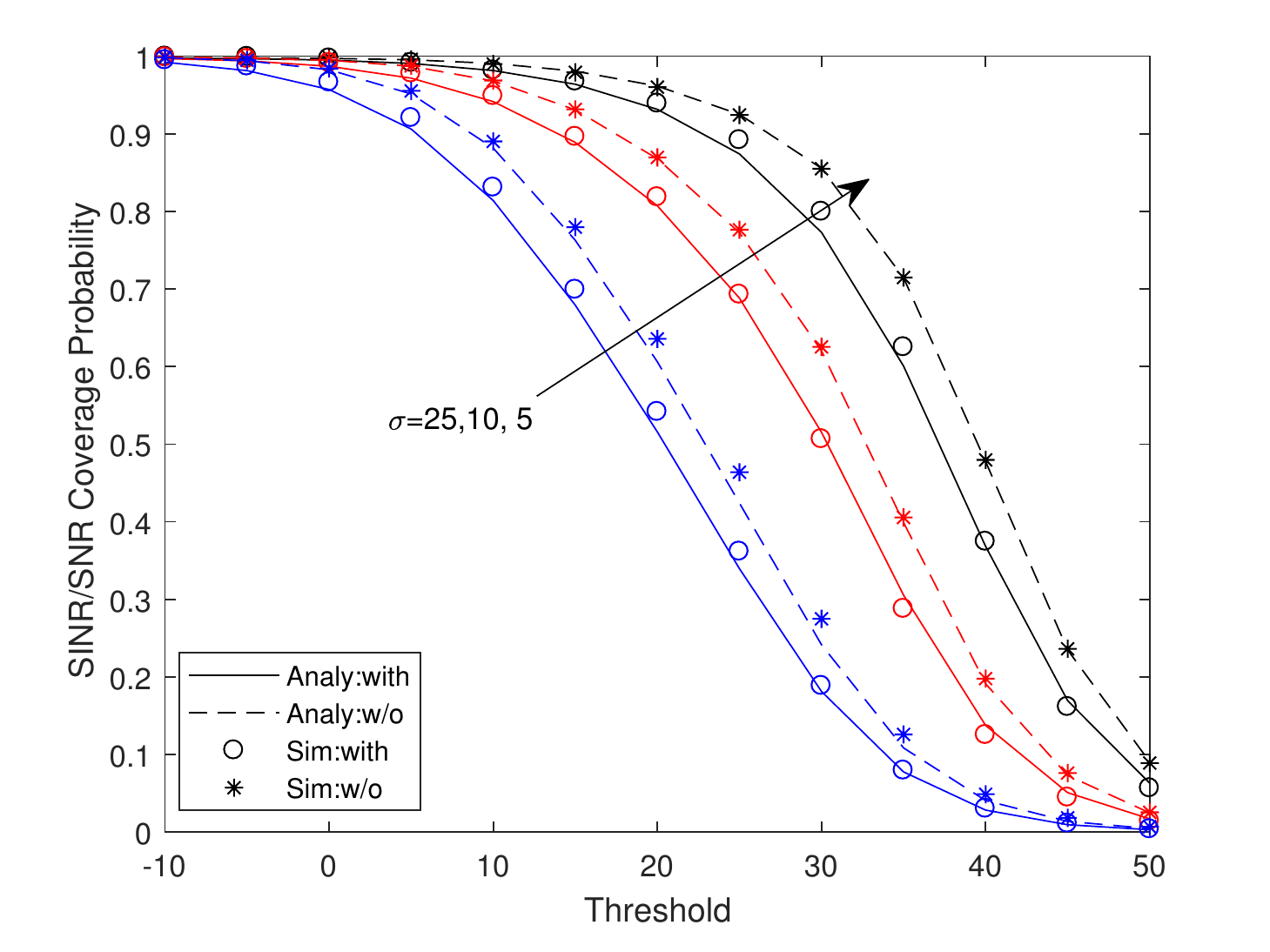} \\
		\subcaption{\scriptsize Power control not considered. }
	\end{minipage}
	\begin{minipage}{0.4\textwidth}
		\centering
		\includegraphics[width=1\textwidth]{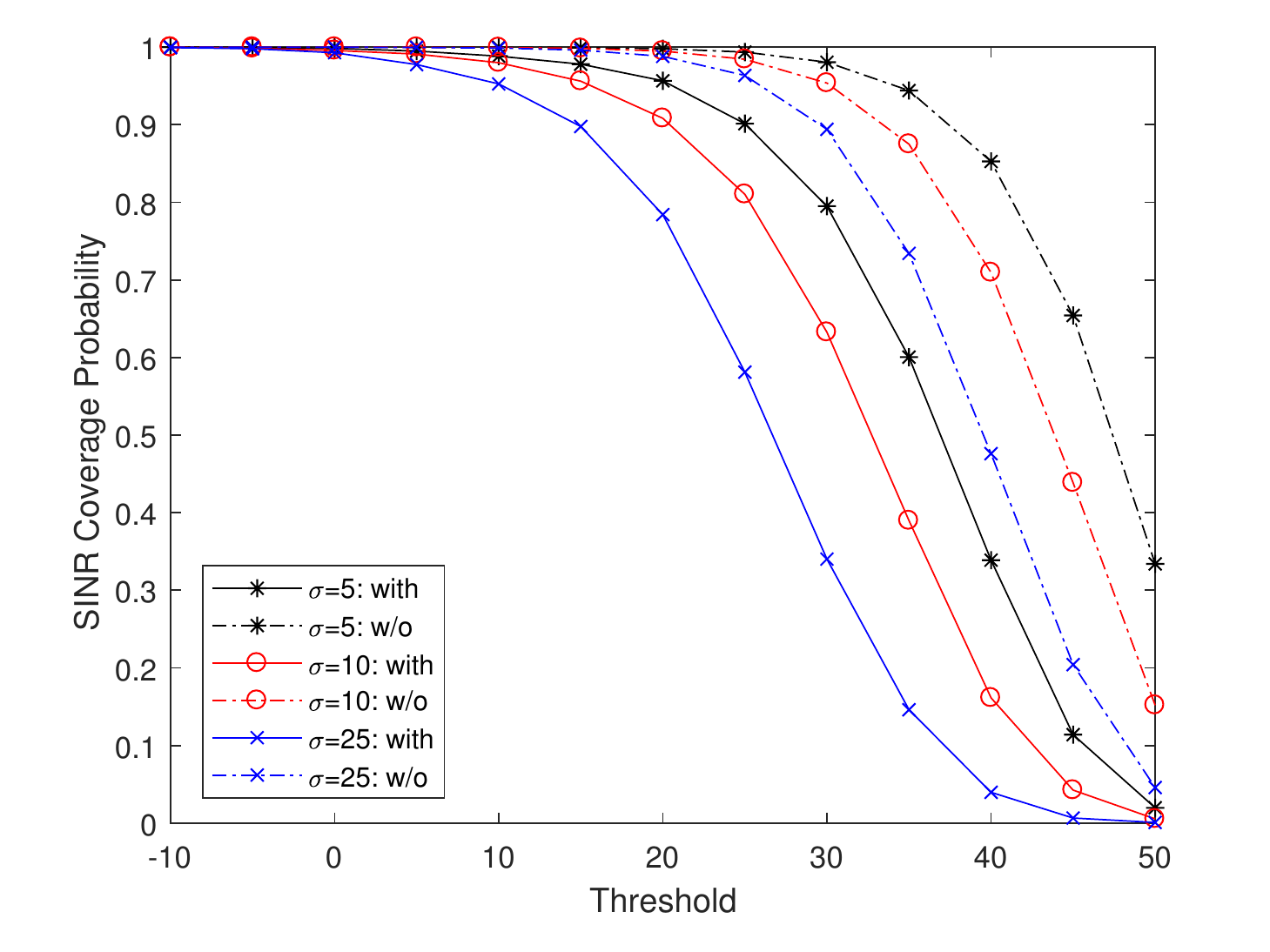}
		\subcaption{\scriptsize Power control considered. }
	\end{minipage}
	\caption{\small SNR and SINR CPs as a function of threshold for different cluster size $\sigma$. \normalsize}
	\label{Figure_interference}
\end{figure}
\subsubsection{Impact of interference}We next investigate the influence of interference in the considered heterogeneous mmWave network. Fig. \ref{Figure_interference} exhibits the total SNR and SINR coverage probabilities of the entire network as a function of the threshold for different values of $\sigma$. While Fig. \ref{Figure_interference}(a) is obtained without any power control (i.e., transmission power is fixed), fractional power control is employed in the performance results in Fig. \ref{Figure_interference}(b). It is obvious that  the interference has a significant impact on the coverage performance as shown in this figure. From Fig. \ref{Figure_interference}(b) we observe that if path loss based fractional power control is used by the UEs, the impact of interference becomes more pronounced.  Note that SINR CPs (described in the legend as ``with" as in ``with interference") are lower than SNR CPs (denoted with ``w/o").

\begin{figure}
	\centering
	\includegraphics[width=0.4 \textwidth]{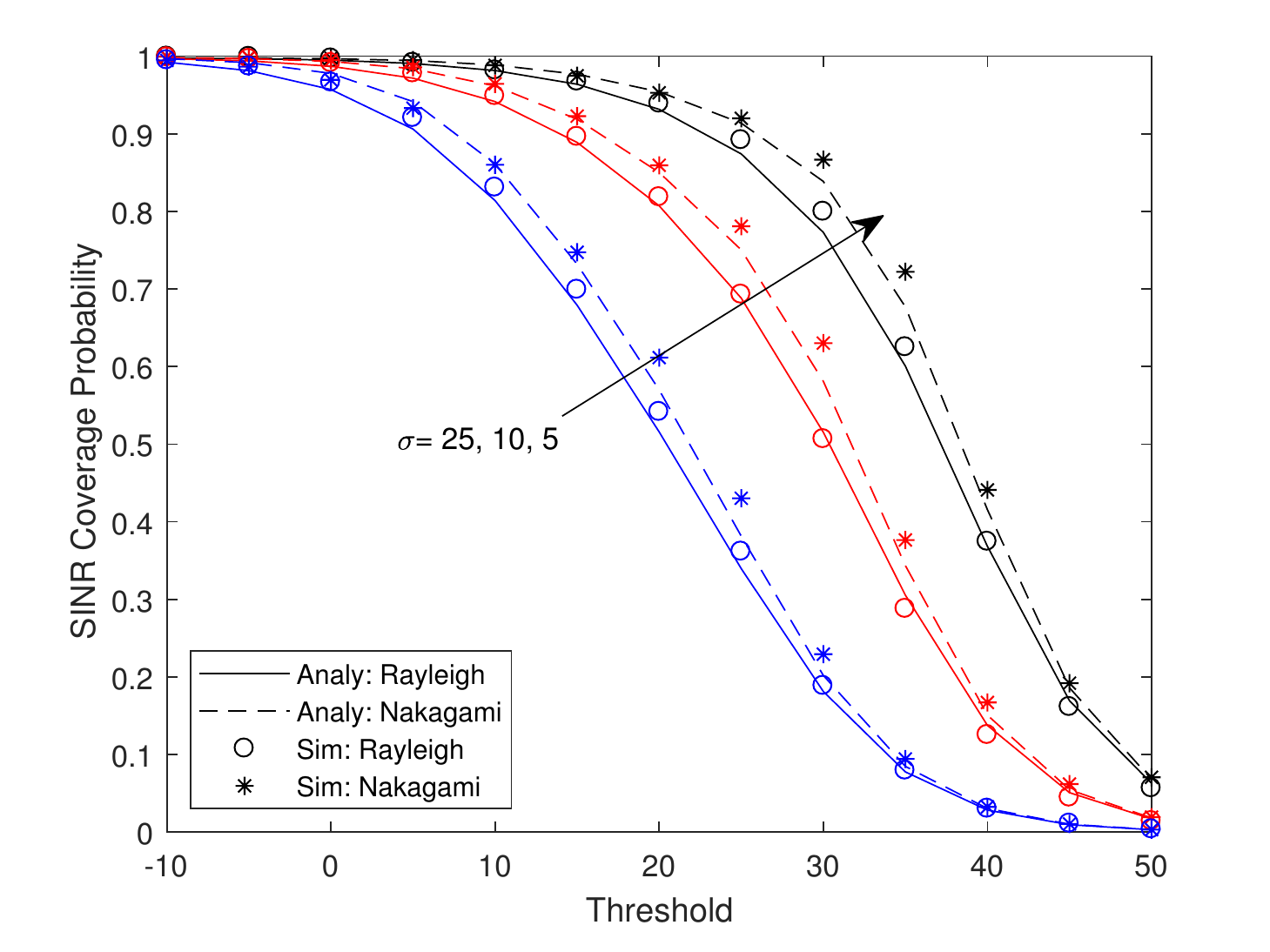}
	\caption{\small SNR and SINR CPs as a function of threshold for different cluster size $\sigma$. \normalsize}
	\label{Figure_fading}
\end{figure}
\subsubsection{Impact of small-scale fading}In Fig. \ref{Figure_fading}, we compare the performances with different types of small-scale fading. In particular, we consider Rayleigh fading and Nakagami fading with parameters $N_L=3$ for LOS and $N_N=2$ for NLOS links.
We observe from the figure that two types of  small-scale fading lead to similar performance trends. However, since Nakagami fading implies more favorable channel conditions, higher CP is achieved with Nakagami fading.

\subsection{Impact of the LOS probability exponent $\epsilon$}
\begin{figure}
	\centering
	\begin{minipage}{0.4\textwidth}
		\centering
		\includegraphics[width=1\textwidth]{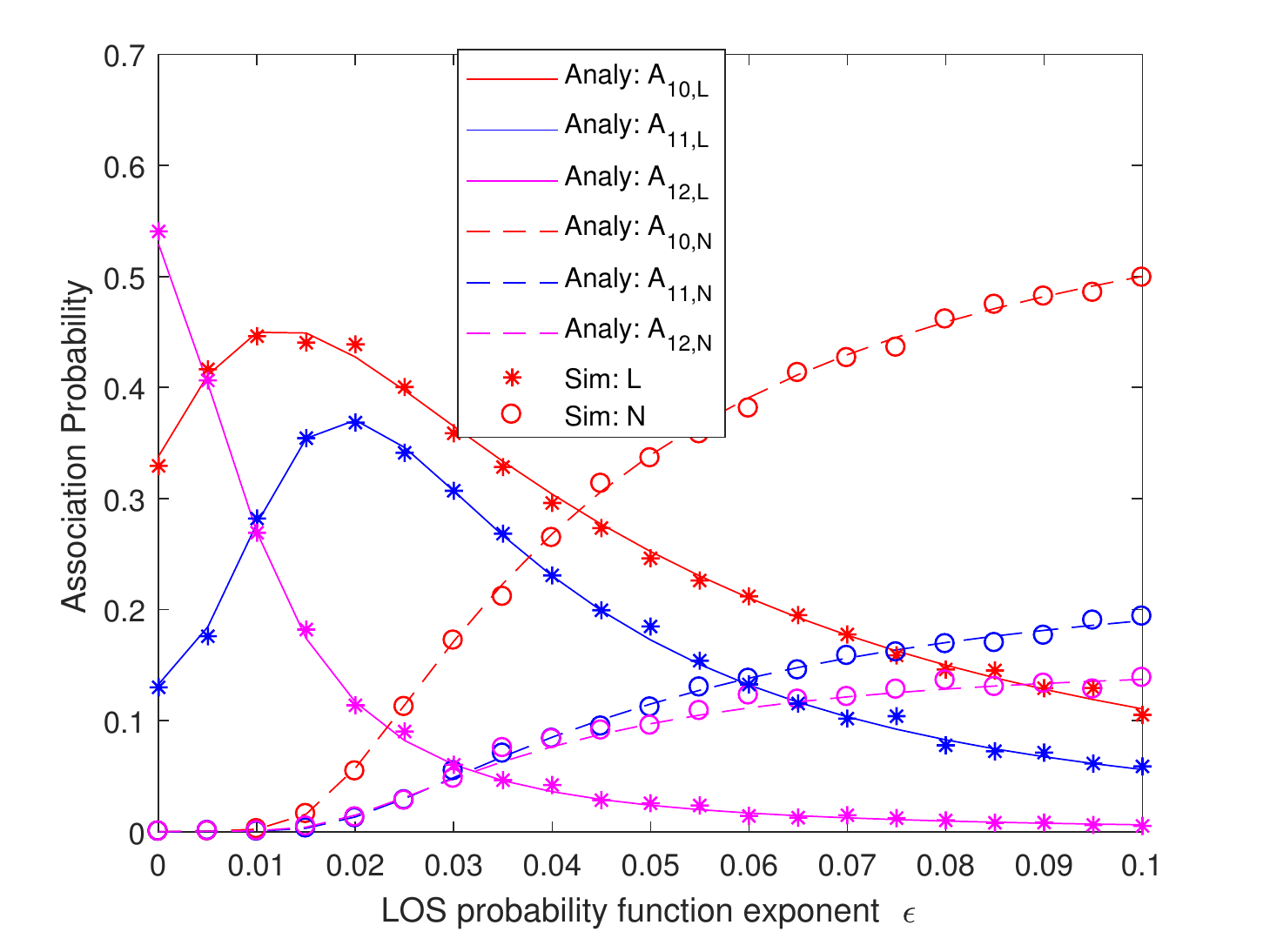} \\
		\subcaption{\scriptsize Association probability. }
	\end{minipage}
	\begin{minipage}{0.4\textwidth}
		\centering
		\includegraphics[width=1\textwidth]{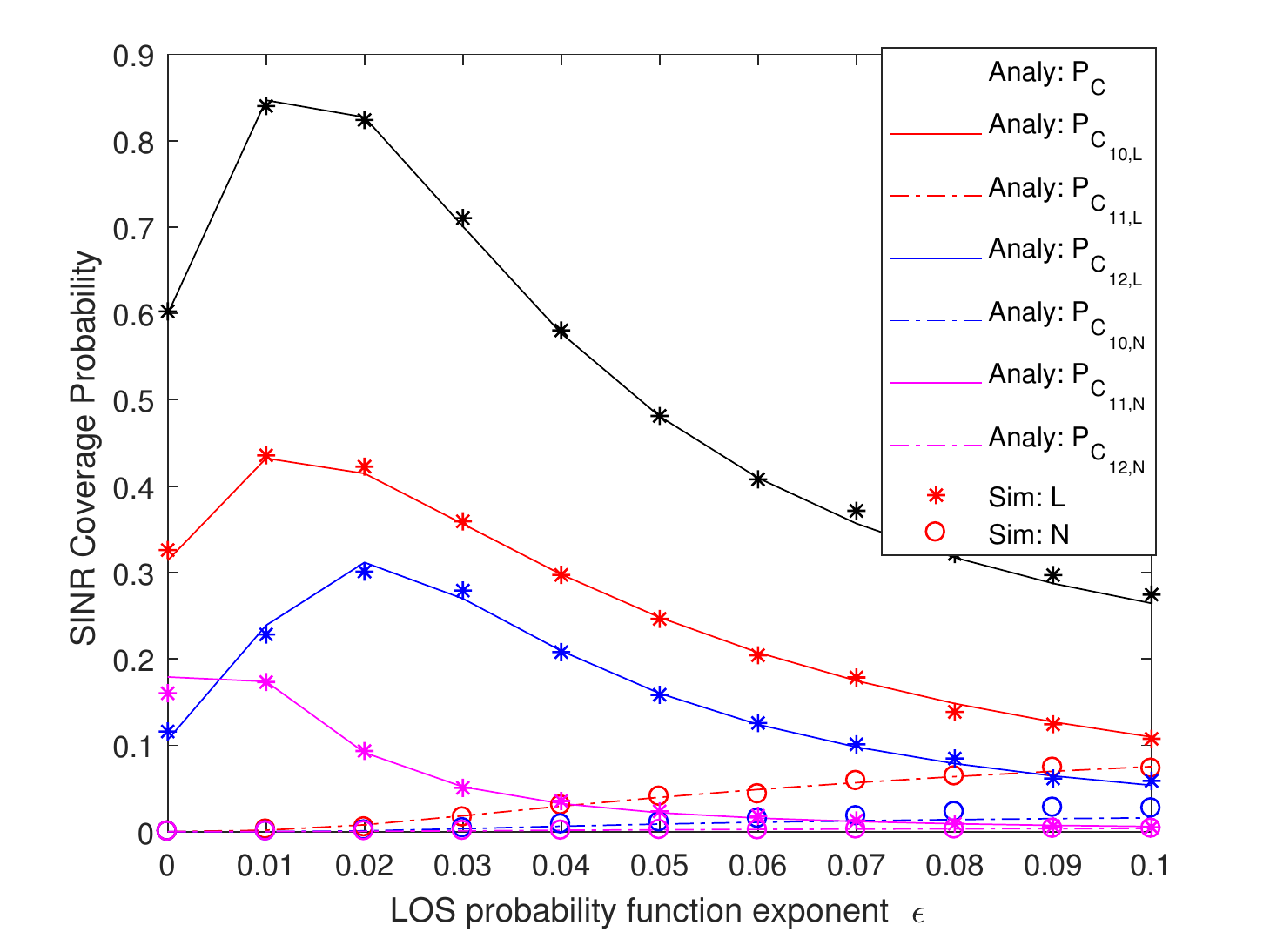}
		\subcaption{\scriptsize  SINR coverage probability.}
	\end{minipage}
	\caption{\small APs and SINR CPs for different tier as a function of the LOS probability function exponent $\epsilon$ when $\sigma=25$ and $T=10$dB. \normalsize}
	\label{AP_CP_e}
\end{figure}
In this subsection, we investigate the impact of the LOS probability exponent $\epsilon$. Note that the smaller $\epsilon$ is, the sparser the environment will be, leading to higher LOS probabilities. And we aim to obtain insights on how the building deployments influence the system performance.
\subsubsection{AP}
Fig. \ref{AP_CP_e}(a) shows the AP as a function of $\epsilon$.  When $\epsilon=0$, all links are LOS and the signals are not attenuated significantly during transmission. Since macrocell BSs provide larger transmit power, we can observe from the figure that $A_{12,L}>A_{10,L}>A_{11,L}$. When $\epsilon$ is increased, more links become NLOS. Thus $A_{12,L}$ decreases dramatically. In the meantime, small-cell BSs have more chance to serve the UEs, and we notice increasing $A_{10,L}$ and $A_{11,L}$ initially. When we increase $\epsilon$ further, more and more links become NLOS. Consequently,  we have increasing $A_{10,N}$, $A_{11,N}$ and $A_{12,N}$, as well as  decreasing $A_{10,L}$, $A_{11,L}$ and $A_{12,L}$. Additionally, when more links are NLOS, signals are attenuated considerably during transmission. And since the distance between UEs and their cluster center BSs are relatively smaller, $A_{10,N}$ increases more than $A_{11,N}$ and $A_{12,N}$.
\subsubsection{CP}
Fig. \ref{AP_CP_e} (b) plots the SINR CP as a function of $\epsilon$. The first observation shown in the figure is that similar to the performance trends with respect to AP, with increasing $\epsilon$, we have increasing $P_{C_{10,N}}$, $P_{C_{11,N}}$ and $P_{C_{12,N}}$. In addition, $P_{C_{10,L}}$ and $P_{C_{11,L}}$ increase initially and then start diminishing, while $P_{C_{12,L}}$ decreases continuously. The second observation is that $P_{C_{10,s}}>P_{C_{11,s}}>P_{C_{12,s}}$ is always satisfied. This is due to the following reasons: 1) no matter which BS the UE is associated with depending on the averaged received power of the UE, in the uplink phase, UE is transmitting information to BSs with the same transmit power; 2) the distance between the UEs and their cluster center BS is generally smaller than to other small-cell BSs and macrocell BSs, leading to the conditional CP of the cluster center being much larger than those of other BSs. The third observation is that the total CP increases at the beginning and then decreases. This is because that $P_{C_{10,L}}$ and $P_{C_{11,L}}$ increase initially, and then the increase in the coverage with NLOS Links cannot compensate the decrease in the coverage with LOS links.

 \subsection{Impact of the biasing factor and transmit power}
\begin{figure}
	\centering
	\begin{minipage}{0.4\textwidth}
		\centering
		\includegraphics[width=1\textwidth]{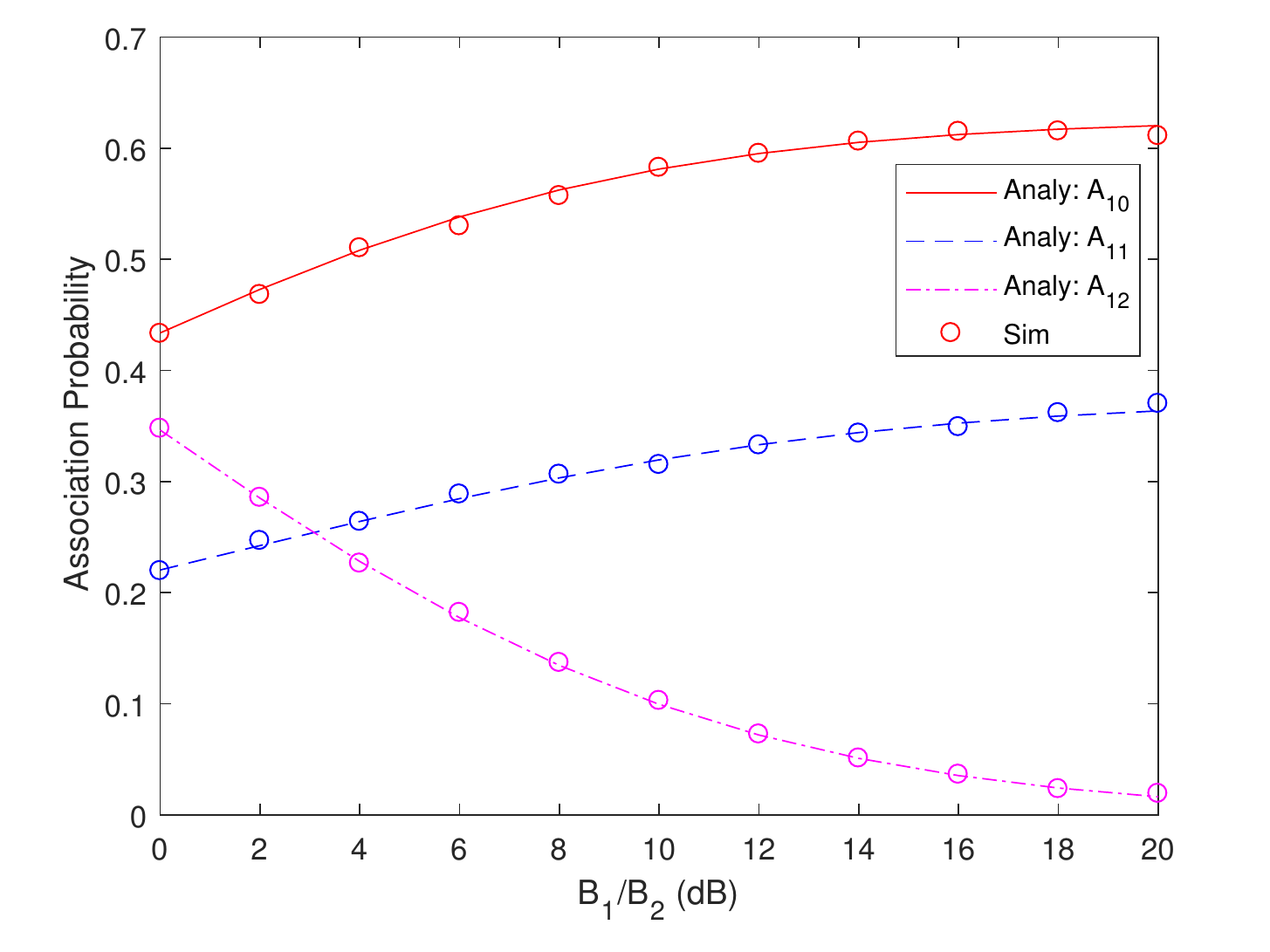} \\
		\subcaption{\scriptsize As a function of $B_1/B_2$. }
	\end{minipage}
	\begin{minipage}{0.4\textwidth}
		\centering
		\includegraphics[width=1\textwidth]{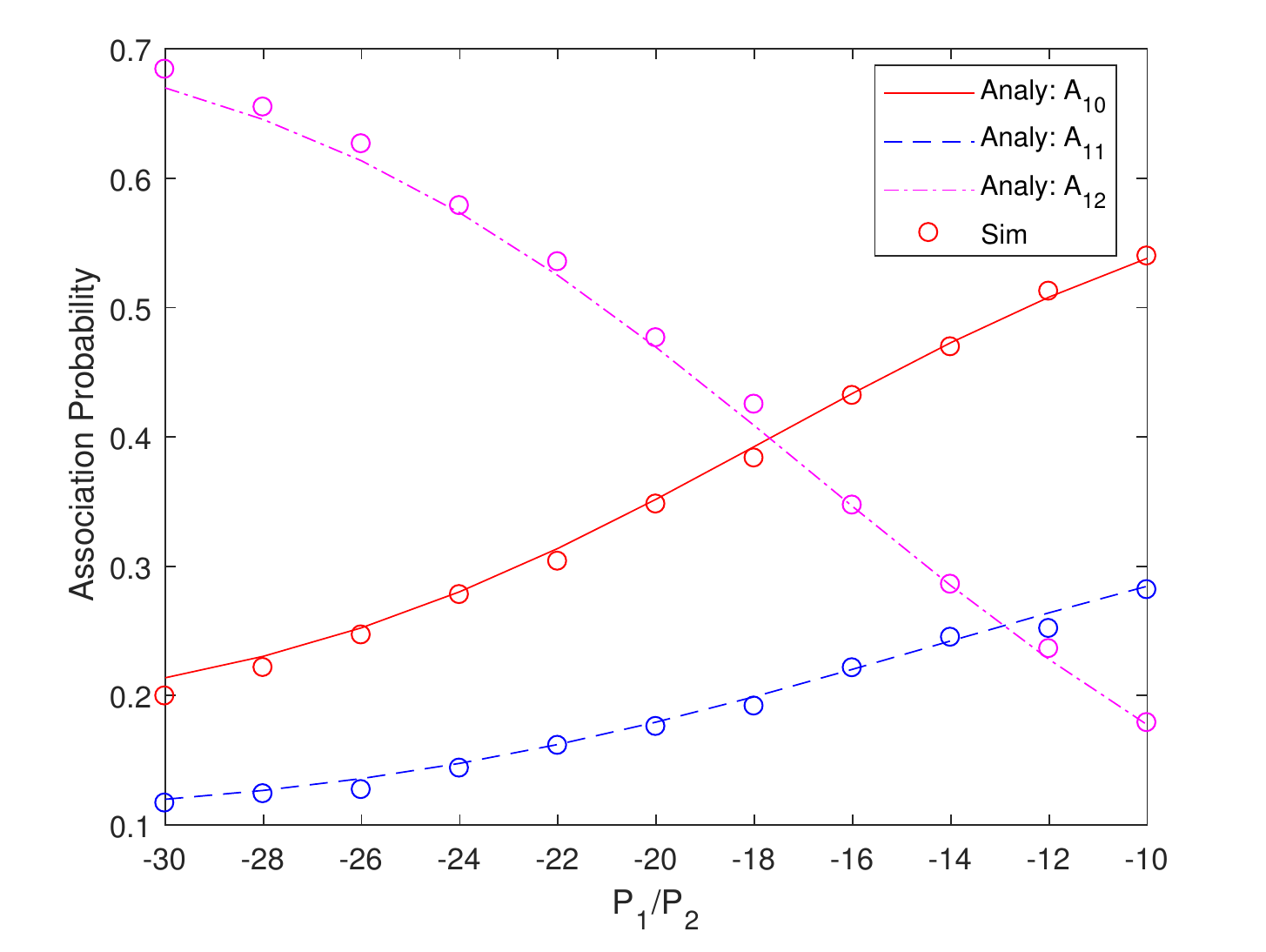}
		\subcaption{\scriptsize As a function of $P_1/P_2$.}
	\end{minipage}
	\caption{\small APs as a function of the biasing factor ratio $B_1/B_2$ and the BS transmit power ratio $P_1/P_2$ when $\sigma=25$ and $T=10$dB. \normalsize}
	\label{AP_Br_Pr}
\end{figure}
\subsubsection{AP}
Fig. \ref{AP_Br_Pr}(a) displays the AP as a function of the biasing factor ratio $B_1/B_2$ in dB, where $B_1$ and $B_2$ are the biasing factors of the small-cell BSs and the macrocell BSs, respectively. When $B_1=B_2$, we have $A_{10}>A_{11}>A_{12}$ since UEs are relatively closer to the cluster centers. On the other hand, when  we fix $B_2=1$ and increase $B_1$, we observe, as expected, that $A_{10}$ and $A_{11}$ grow, while $A_{12}$  diminishes.

Fig. \ref{AP_Br_Pr}(b) shows the AP as a function of the transmit power ratio $P_1/P_2$ in dB. At the beginning, since $P_2$ is much larger than $P_1$, even though the distance between the macrocell  BSs and UEs are large, the macrocell BSs still have a high probability to connect with the UEs. When we fix $P_2$ and increase $P_1$, the small-cell BSs start having larger transmit power, and as a result  $A_{10}$ and $A_{11}$ increase.
\subsubsection{CP}
\begin{figure}
	\centering
	\begin{minipage}{0.4\textwidth}
		\centering
		\includegraphics[width=1\textwidth]{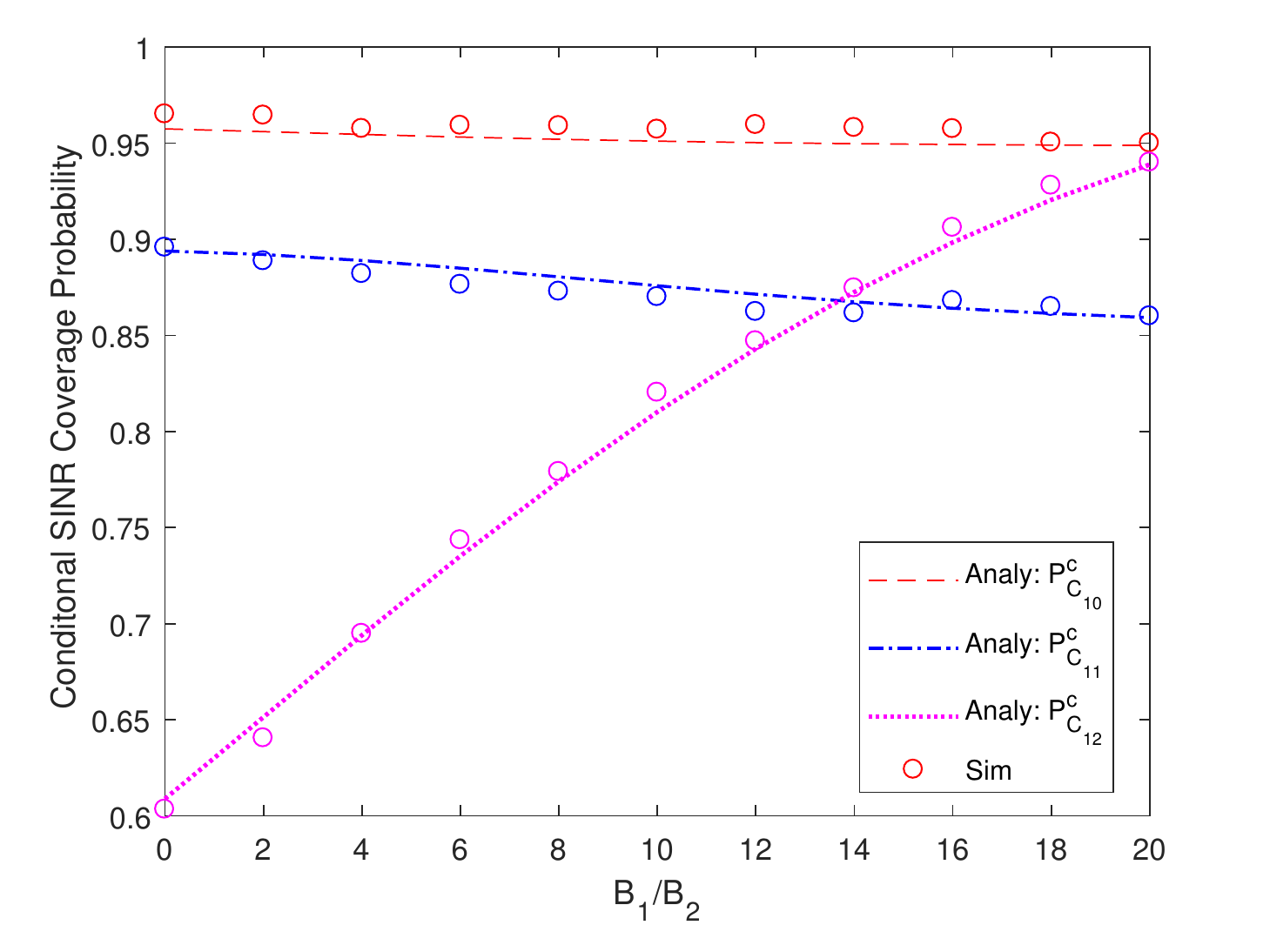} \\
		\subcaption{\scriptsize Conditional SINR coverage probabiliy. }
	\end{minipage}
	\begin{minipage}{0.4\textwidth}
		\centering
		\includegraphics[width=1\textwidth]{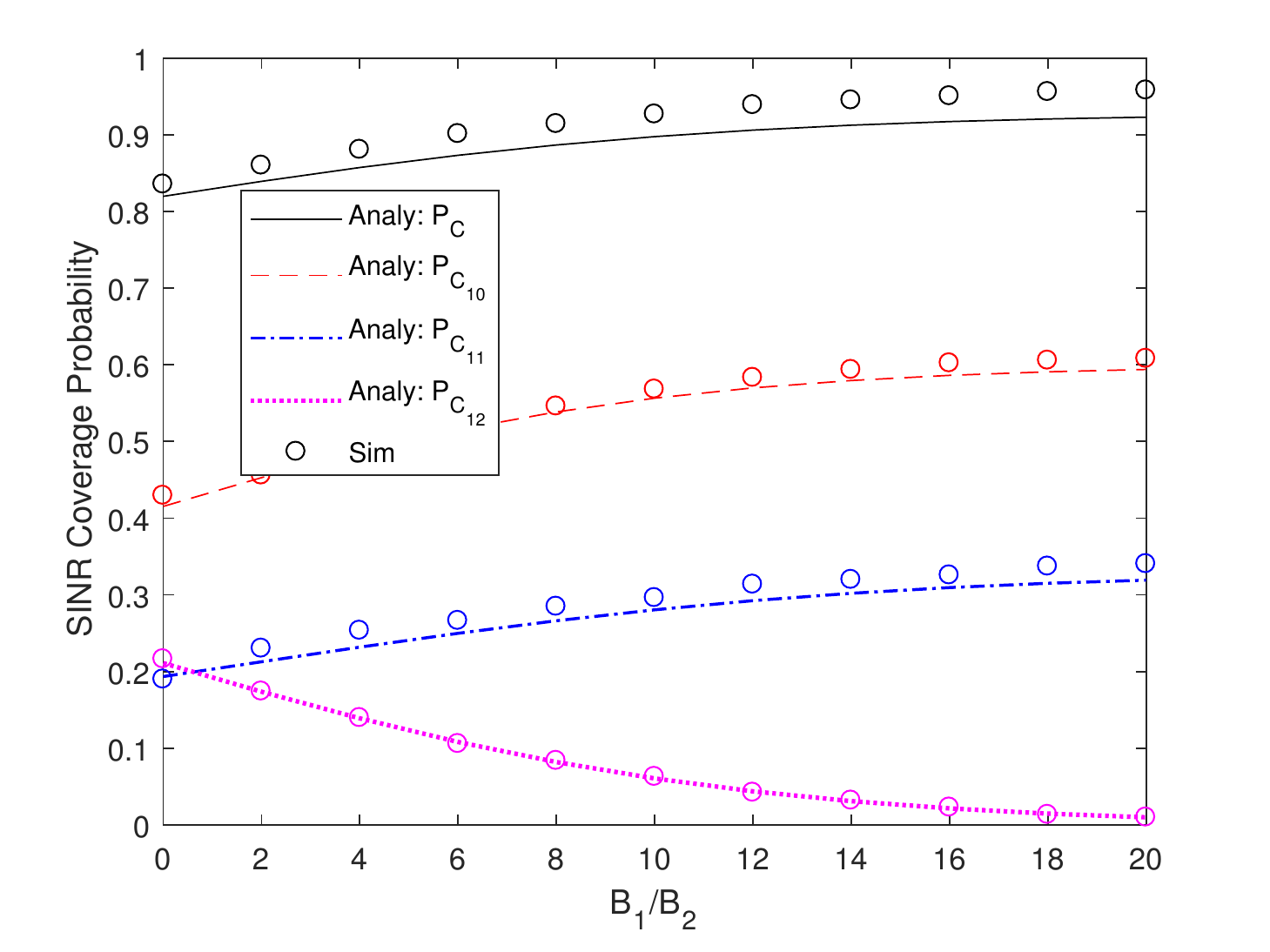}
		\subcaption{\scriptsize SINR coverage probabiliy. }
	\end{minipage}
	\caption{\small SINR CP as a function of $B_1/B_2$ when $\sigma=25$ and $T=10$dB. \normalsize}
	\label{CP_Br}
\end{figure}

Due to their impact on the association probabilities, biasing factor ratio and the BS transmit power ratio influence the choice of the associated reference BS, and consequently affect the SINR coverage probability of each reference BS. Fig. \ref{CP_Br} plots the conditional SINR coverage probability ($P^c_{C_{ij}}$) of each tier of reference BSs, and the SINR coverage probability of each tier ($P_{C_{ij}} = P^c_{C_{ij}} A_{ij}$)). From Fig. \ref{CP_Br}(a), we can observe that the conditional SINR coverage probability of the cluster center BSs $P^c_{C_{10,s}}$ and other small-cell BSs $P^c_{C_{11,s}}$ decrease, while that of the macrocell BSs $P^c_{C_{12,s}}$ increases with increasing $B_1/B_2$. The reason is that, as $B_1/B_2$ gets large, UEs increasingly prefer to associate with small cell BSs, and consequently the small-cell BSs start having larger association priority, while the macrocell BSs have small association priority. In this case, if the typical UE is still associated with a macrocell BS, then that link should be in really good condition, and hence has large coverage probability. Therefore, with increasing $B_1/B_2$, given that the typical UE is associated to the macrocell BS, the macrocell BS should provide strong power and have larger coverage probability, leading to increasing $P^c_{C_{12,s}}$. Conversely, for large $B_1/B_2$, small-cell BSs, which do not need large power to get associated to UEs, have smaller coverage.

 On the other hand, due to the variations in the association probability of each tier shown in Fig. \ref{AP_Br_Pr}(a), we have the performance in Fig. \ref{CP_Br}(b). In particular, we note that $P_{C_{10}}$ and $P_{C_{11}}$ both increase, because $A_{10}$ and $A_{11}$ grow significantly with increasing $B_1/B_2$.
Since the increment in $P_{C_{10}}$ and $P_{C_{11}}$ is larger than the decrease in $P_{C_{12}}$, we have a growing total network CP.. Therefore, total SINR CP of the network is an increasing function of $B_1/B_2$. Regarding the transmit power, if we increase the transmit power of small-cell BSs, the total CP of the network will also increase.


\subsection{Impact of the density of BSs}
\begin{figure}
	\centering
	\includegraphics[width=0.4 \textwidth]{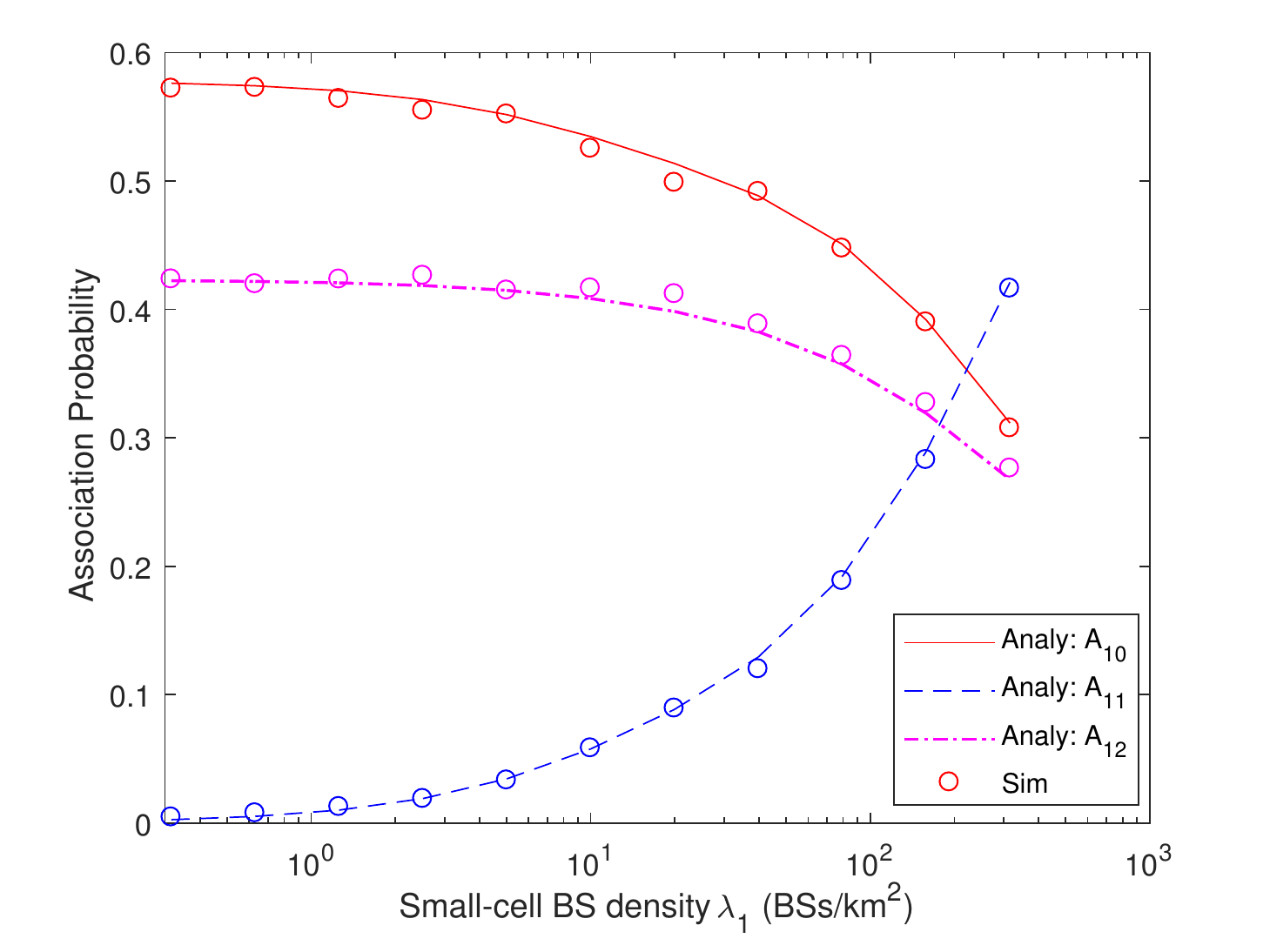}
	\caption{\small APs as a function of the density of small-cell BSs $\lambda_1$, when $\sigma=25$ and $T=10$dB.  \normalsize}
	\label{AP_Lr}
\end{figure}
\subsubsection{AP}
Fig. \ref{AP_Lr} shows the AP as a function of $\lambda_1$ when $\lambda_2$ = 10 / km$^2$ and $\sigma_u=25$. Initially, the density of small-cell BSs is small, UEs are still clustered around them, and the macrocell BSs are not densely distributed. Hence, the UEs are likely to be served by their cluster center BSs, i.e. $A_{10}>A_{12}>A_{11}$. When we increase $\lambda_1$, more and more small-cell BSs are deployed. The distance between the UEs and their cluster center BS does not change, while the distance from the UEs to other small-cell BSs becomes smaller. Hence, the USs become more and more likely to be served by other small-cell BSs. In other word, $A_{11}$ increases, while $A_{10}$ and $A_{12}$ both decrease.
\subsubsection{CP}
\begin{figure}
	\centering
	\begin{minipage}{0.4\textwidth}
		\centering
		\includegraphics[width=1\textwidth]{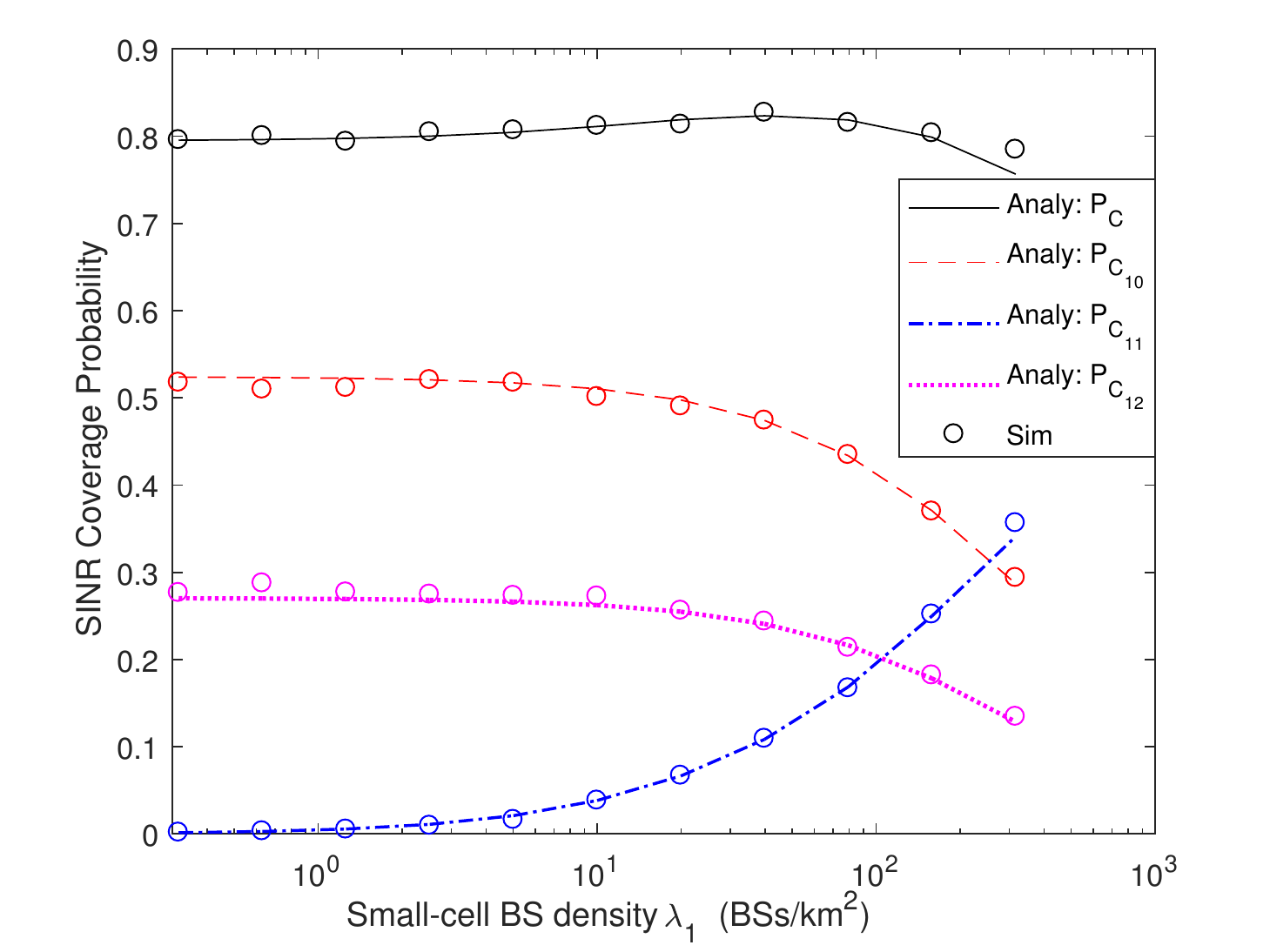} \\
		\subcaption{\scriptsize  }
	\end{minipage}
	\begin{minipage}{0.4\textwidth}
		\centering
		\includegraphics[width=1\textwidth]{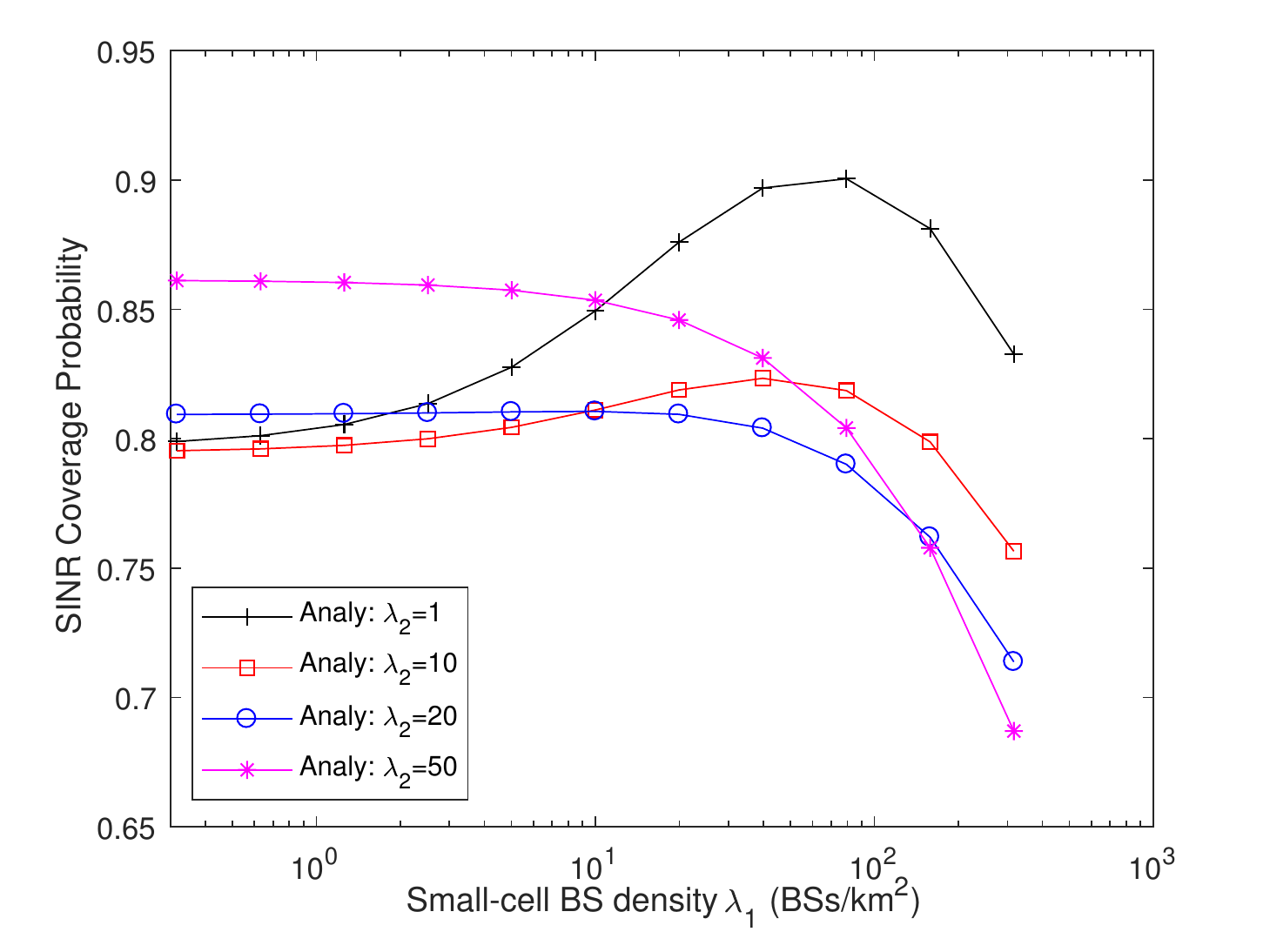}
		\subcaption{\scriptsize  }
	\end{minipage}
	\caption{\small SINR CPs of each tier and the entire network as a function of the density of small-cell BSs $\lambda_1$, when $\sigma=25$ and $T=10$dB. \normalsize}
	\label{CP_Lr}
\end{figure}
Fig. \ref{CP_Lr}(a) shows the SINR CP as a function of $\lambda_1$. Via the CP of each tier, we can observe that with an increasing $\lambda_1$, $P_{C_{10}}$ and $P_{C_{12}}$ diminish, while $P_{C_{11}}$ grows. Initially, the increase in $P_{C_{11}}$ is larger than the decrease in $P_{C_{10}}$ and $P_{C_{12}}$, leading to the result that total CP increases. As we increase $\lambda_1$ further since the interference grows, the increase in $P_{C_{11}}$ cannot compensate  the decrease in $P_{C_{10}}$ and $P_{C_{12}}$, and consequently the total CP decreases.

Fig. \ref{CP_Lr}(b) plots the SINR CP vs. small-cell BS density $\lambda_1$ for different densities $\lambda_2$ of the macrocell BSs. From the figure, we observe that when $\lambda_1=1$ BS/km$^2$ which is small, increasing $\lambda_2$ improves the total network coverage. This is expected since we have a sparse network and interference can be ignored, and increasing the number BSs results in network performance improvement. On the other hand, when $\lambda_1=100$ BSs/km$^2$, increasing $\lambda_2$ lowers the SINR CP performance. In this case, the network have more small-cell BSs, including the cluster center BSs and other small-cell BSs,  which together can provide good coverage. Increasing the macrocell BS density will lead to more interference, and also will result in UEs getting connected to macrocell BSs with potentially larger link distances.  Thus, in this case, the total SINR coverage generally diminishes when $\lambda_2$ is increased.

These two figures give us the following useful insights: 1) when the macrocell BSs are sparsely deployed (e.g., $\lambda_2 \leq 10$ BSs/km$^2$), there is an optimal density of small-cell BSs to maximize the uplink SINR CP of the entire network; while when the macrocell BSs are densely deployed, increasing the small-cell BS density will lead to smaller uplink SINR CP; 2) when the small-cell BSs are sparsely deployed, increasing the density of macro BS will improve the uplink CP; while the opposite will happen for densely deployed small-cell BS networks.

\section{Conclusion} \label{Con}

We have studied a $K$-tier heterogeneous mmWave uplink cellular network with clustered UEs. In particular, the correlation between the locations of the user UEs and BSs is characterized according to a Gaussian distribution, leading to the Thomas cluster processes. Specific and practical LOS and NLOS models are adopted with different parameters for different tiers. We have first characterized the PDFs and CCDFs of different distances from BSs to UEs. Then, we have considered the coupled association strategy, meaning that the UEs are associated with the same BS for both downlink and uplink. Largest long-term averaged biased received power criterion is considered in this paper, and general expressions for association probabilities of different BSs to different UEs are also provided. Following the identification of the association probabilities, we have characterized the Laplace transforms of the inter-cell interference and intra-cluster interference. Using tools from stochastic geometry, we have provided general expressions of the SINR coverage probability in each tier. Additionally, we have extended our work to the Nakagami fading model and the case in which fractional power control is adopted by the UEs. Moreover, we have addressed several special cases, e.g., the noise-limited case, interference-limited case, and one-tier model. Finally, we have addressed the average ergodic spectral efficiency.
Via numerical and simulation results, we have confirmed the analytical characterizations and the derived expressions, and investigated the impact of important system parameters. For instance, we have observed that the interference has a noticeable influence on the coverage performance and the type of fading has a certain impact on the SINR coverage performance but Nakagami and Rayleigh fading lead to similar performance trends. Coverage probability can be improved by decreasing the cluster size, increasing the biasing factor and the transmit power of the small-cell BSs. The densities of BSs also affect the system performance. 

\appendix

\subsection{Proof of Theorem 1} \label{Proof_CP}
Given $S_{ij,s}$, the SINR coverage probability can be obtained as follows:
\begin{align}
& P^c_{C_{ij,s}} \overset{}{=} \prob (\tsinr_{ij,s} > T_j | S_{ij,s}) \notag \\
&\overset{(a)}{=}\prob \left(\frac{P_u G_0 h_j \kappa^{-1}_s r_{ij,s}^{-\alpha_s}}{\sigma_{n}^2+I_{j0}+\sum\limits_{k=1}^{K_u} I_{jk}} > T_j \right) \notag  \\
&=\prob \left( h_j > \frac{T_j \kappa_s r_{ij,s}^{\alpha_s}}{P_u G_0 } \left(\sigma_{n}^2+ I_{j0}+\sum_{k=1}^{K_u} I_{jk} \right)  \right) \notag  \\
&\overset{(b)}{=}\E_{r_{ij,s}} \left[ \exp\left( - \frac{T_j \kappa_s r_{ij,s}^{\alpha_s}}{P_u G_0 } \left(\sigma_{n}^2+ 	I_{j0}+\sum_{k=1}^{K_u} I_{jk} \right) \right) \right] \notag \\
&\overset{(c)}{=} \E_{r_{ij,s}} \left[ e^{-\mu^s_{ij} \sigma^2_{n }} \cL_{I_{i0}} (\mu^s_{ij}) \prod\limits_{k=1}^{K_u} \cL_{I_{ik}} (\mu^s_{ij})  \right] ,
\end{align}
where $\mu_{ij}^s= \frac{T_i \kappa_s r_{ij,s}^{\alpha_s}}{P_u G_0} $, and  $\cL_{I}(\mu)= \E \texp ( - \mu  I  )$ is the Laplace transform of $I $ evaluated at $\mu$. (a) follows from the expression of SINR given $S_{ij,s}$ is true. (b) follows from the moment generating function (MGF) of $h \sim \exp(1)$. (c) is due to the fact that the noise component and the interference from each tier are independent of each other. Note that when $I_{j0}=0$, $\cL_{I_{i0}} (\mu^s_{ij}) =1$.


\subsection{Proof of Theorem 2}\label{Proof_Laplace}
1) For $k\in \mK$, we have
$\cL_{I_{ik}}(\mu_{ij}^s) = \prod_G \prod_{a \in \{L, N \} }  \cL_{I_{ik}^{Ga}}(\mu_{ij}^s)$.
Using $\x_{jk,n}$ to denote the vector from the reference BS to the $n^{th}$  BS in the $k^{th}$ tier, and $\y_{k0}$ to denote the vector from the $k^{th}$ tier BS to its active cluster member UE, we get $r_{jk,n} = ||\x_{jk,n}+\y_{k0}||$.
\begin{align}\label{APP:eq:Lap}
    &\cL_{I_{jk}^{Ga}}(\mu_{ij}^s) = \E [\texp ( - \mu_{ij}^s I_{jk}^{Ga})] \notag \\
    &\overset{(a)}{=}  \E \left[\texp \left(-\mu_{ij}^s \sum_{n\in\Phi'_{uk}} P_u G h_k \kappa^{-1}_a r_{jk,n}^{-\alpha_a} \right) \right] \notag \\
    &= \E \left[ \prod_{n\in\Phi'_{uk}}  \E_{h_k} \left[\texp(-\mu_{ij}^s P_u G h_k \kappa^{-1}_a r_{jk,n}^{-\alpha_a})  \right]\right ] \notag \\
    &\overset{(b)}{=}  \E \left[ \prod_{n\in\Phi'_{uk}}   \frac{1}{1+ \mu_{ij}^s P_u G \kappa^{-1}_a r_{jk,n}^{-\alpha_a}}  \right] \notag\\
    &\overset{}{=}  \E \left[ \prod_{n\in\Phi'_{k}}   \frac{1}{1+ \mu_{ij}^s P_u G \kappa^{-1}_a ||\x_{jk,n}+\y_{k0}||^{-\alpha_a}}  \right] \notag  \\
    &\overset{}{=}  \E_{\x_{jk,n}} \left[ \prod_{n\in\Phi'_{k}}   \E_{\y_{k0}} \left[ \frac{1}{1+ \mu_{ij}^s P_u G \kappa^{-1}_a ||\x_{ik,n}+\y_{k0}||^{-\alpha_a}} \right]  \right] \notag  \\
    &\overset{(c)}{=}  e^{  -  \int_{\R^2} \lambda'_{k}  \left( 1- \E_{\y_{k0}}  \left[ \frac{1}{1+ \mu_{ij}^s P_u G \kappa^{-1}_a ||\x_{jk,n}+\y_{k0}||^{-\alpha_a}}  \right]  \right) d\x_{jk,n} } \notag \\
    &\overset{(d)}{=} e^ {  - \int_{\R^2}  \lambda'_{k} \E_{\y_{k0}}  \left[ \frac{1}{1+( \mu_{ij}^s P_u G \kappa^{-1}_a ||\x_{jk,n}+\y_{k0}||^{-\alpha_a})^{-1}} \right]  d\x_{ik,n}} \notag\\
    &\overset{(e)}{=} e^{  - \int_{\R^2}  \int_{\R^2}  \lambda'_{k}  \frac{1}{1+( \mu_{ij}^s P_u G \kappa^{-1}_a ||\x_{jk,n}+\y_{k0}||^{-\alpha_a})^{-1}}  f_Y(\y_{k0}) d\y_{k0}  d\x_{jk,n} } \notag\\
    &\overset{(f)}{=} e^{  -2 \pi  \int_0^{\infty} \int_0^{\infty} \lambda'_{k}    \frac{f_R(r_{jk,n}|v_{jk,n})  }{1+( \mu_{ij}^s P_u G \kappa^{-1}_a r_{jk,n}^{-\alpha_a})^{-1}}  dr_{jk,n} v_{jk,n} dv_{jk,n} } \notag\\
    &\overset{(g)}{=} e^{ -2 \pi \int_0^{\infty} \lambda_k p_G  p^a_{\alpha k}(r_{jk,n})\left( \frac{1}{1+( \mu_{ij}^s P_u G \kappa^{-1}_a r_{jk,n}^{-\alpha_a})^{-1}} \right) r_{jk,n} dr_{jk,n} }
\end{align}
where $\Phi'_{uk}$ is the union of interfering UEs in the $k^{th}$ tier, $\Phi'_k$ is the thinned union of BSs in the $k^{th}$ tier due to $p_G p^s_{\alpha k}(r_{jk,n})$, and $\lambda'_{k}=p_G p^s_{\alpha k}(r_{jk,n}) \lambda_k$. By the definition of interference, we obtain (a). (b) follows from the fact that $h \sim \exp(1)$. (c) is due to the computation of the probability generating function (PGFL) of PPP, which describes the distribution of BSs, since we compute the expectation with respect to $x_{jk,n}$ in this step. (d) is determined from the fact that $1- \E[\frac{1}{1+x}]=\E[\frac{x}{1+x}]=\E[\frac{1}{1+x^{-1}}]$. (e) follows by plugging in the expression of the expected value with respect to $y_{k0}$. (f) is obtained by converting the coordinates from Cartesian to polar, and $v_{jk,n} = ||\x_{jk,n} ||$. (g) follows from an approximation based on the property of the Rician distribution that $\int_0^\infty f_R(r|v) v dv= r$.

2) For $k=0$, we have
    \begin{align}\label{Lapi0}
       \cL_{I_{j0}}(\mu_{ij}^s)=\sum_G \sum_{a\in \{L,N\}} p_G D^a_{j0} \cL_{I_{j0}^{Ga}}(\mu_{ij}^s)
    \end{align}
    and since only one UE in the cluster inflicts interference, which can be via either a LOS or NLOS link, and the antenna gain is selected from $\{MM, Mm, mm  \}$, we can express
 \begin{align}
    &\cL_{I_{j0}^{Ga}}(\mu_{ij}^s)= \E \texp (-\mu_{ij}^s I_{j0}^{Ga} ) \notag \\
    &=\E \texp \left[ -\mu_{ij}^s P_u G h \kappa^{-1}_a r_{j0}^{-\alpha_a} \right] \notag \\
    &\overset{(a)}{=} \int_0^\infty \frac{1}{1+ \mu_{ij}^s P_u G \kappa^{-1}_a r_{j0}^{-\alpha_a}} f_{R^s_{j0}}(r_{j0}) dr_{j0}
 \end{align}
where (a) is obtained by using the MGF of $h \sim \exp(1)$.

\subsection{Proof of Lemma 3}\label{Proof_PowerControl}
1) For the interfering cluster member UEs $k=0$, we have
\begin{align}
	&\cL_{I_{j0mb}^{Ga}}(\mu_{ij}^s)= \E \texp (-\mu_{ij}^s I_{j0}^{Ga} ) \notag \\
	&=\E \texp \left[ -\mu_{ij}^s (\kappa_b t^{\alpha_b})^{\tau} G h \kappa^{-1}_a r_{j0}^{-\alpha_a} \right] \notag \\
	&\overset{(a)}{=} \int_0^\infty \int_0^\infty \frac{\hat{f}_{r_{km,b} } (t )f_{R^s_{j0}}(r_{j0})}{1+ \mu_{ij}^s (\kappa_b t^{\alpha_b})^{\tau} G \kappa^{-1}_a r_{j0}^{-\alpha_a}}  dt dr_{j0}
\end{align}
where (a) is obtained by computing the expectation with respect to $t$ and applying the  MGF of $h$.

2) For $k\in K_u$, we have

	\begin{align}
	&\cL_{I_{jkmb}^{Ga}}(\mu_{ij}^s) = \E [\texp ( - \mu_{ij}^s I_{jkmb}^{Ga})] \notag \\
	&\overset{}{=}  e^{  -  \int_{\R^2} \lambda'_{k}  \left( 1-  \E_{\y_{k0}} \E_t  \left[ \frac{1}{1+ \mu_{ij}^s P_u(t) G \kappa^{-1}_a ||\x_{jk,n}+\y_{k0}||^{-\alpha_a}}  \right]  \right) d\x_{jk,n} } \notag \\
	&\overset{(a)}{=} e^{ -2 \pi \int_0^{\infty} \lambda_k p_G  p^a_{\alpha k}(r_{jk,n}) \E_t\left[ \frac{1}{1+( \mu_{ij}^s P_u(t) G \kappa^{-1}_a r_{jk,n}^{-\alpha_a})^{-1}} \right] r_{jk,n} dr_{jk,n} } \notag\\
	&\overset{(b)}{=} e^{ -2 \pi \int_0^{\infty} \lambda_k p_G  p^a_{\alpha k}(r_{jk,n})  \int_{0}^{\infty} \frac{A_{km,b} \hat{f}_{r_{km,b} (t) dt}}{1+( \mu_{ij}^s P_u(t) G \kappa^{-1}_a r_{jk,n}^{-\alpha_a})^{-1}}  r_{jk,n} dr_{jk,n} }  \notag\\
	&\overset{}{=} e^{ -2 \pi A_{km,b}  \lambda_k p_G   \int_0^{\infty} \int_{0}^{\infty}    \frac{p^a_{\alpha k}(r_{jk,n}) \hat{f}_{r_{km,b} (t) r_{jk,n} }}{1+( \mu_{ij}^s (\kappa_b t^{\alpha_b})^{\tau} G \kappa^{-1}_a r_{jk,n}^{-\alpha_a})^{-1}} dt  dr_{jk,n} }
	\end{align}
where $t$ is the distance from the interfering UE to its own associated BS. Following the same derivation in (\ref{APP:eq:Lap}) we obtain (a). In (b), we get the expectation with respect to $t$.

\subsection{Proof of Remark 1}\label{Proof_NoiseLimited}
Given the special case, we have
\begin{align}
	&A_{10} = \E_{r_{i0}} \left[\Fbar_{r_1}(r) \Fbar_{r_2}\left( \left(\frac{P_2 B_2}{P_1 B_1} \right)^{ \frac{1}{2} } r\right)  \right] \notag \\
	&= \int_{0}^{\infty} e^{\pi \lambda_1 r^2}  e^{\pi \lambda_1 r^2 \frac{P_2 B_2}{P_1 B_1}}  \frac{r}{\sigma^2} e^{-\frac{r^2}{2 \sigma^2}} dr \notag \\
	& = \frac{1}{\sigma^2} \int_{0}^{\infty} r e^{- (\pi \lambda_1 + \pi \lambda_1  \frac{P_2 B_2}{P_1 B_1} +\frac{1}{2 \sigma^2}) r^2} dr
	 = \frac{1}{\sigma^2} \int_{0}^{\infty} r e^{- C_0 r^2} dr \notag \\
	 & =  \frac{1}{\sigma^2} \left(-\frac{1}{2C_0} e^{- C_0 r^2} \right) \bigg|^{\infty}_0 =  \frac{1}{2C_0 \sigma^2}
\end{align}
Following a similar approach, we can obtain $A_{11}$ and $A_{12}$. Then, we can have the PDFs of the conditional distances as $\hat{f} _{r{10}} (x) = \frac{x}{\sigma^2} e^{-C_0 x^2} /A_{10}$, $\hat{f} _{r{11}} (x) = 2\pi \lambda_1 x e^{-C_0 x^2}/ A_{11}$, and $\hat{f} _{r{12}} (x) = 2\pi \lambda_2 x e^{-C_2 x^2}/ A_{12}$, where $C_0$ and $C_2$ are given in Lemma 4. Using the PDFs, we can be calculate the SNR coverage probability as
\begin{align}
  P^c_{C_{10}} &= \E_{r_10} \left[e ^{-\mu \sigma_n^2}\right]
  = \int_{0}^{\infty} e^{-\frac{T \sigma_n^2}{P_u G_0}r^2} \frac{r}{\sigma^2} e^{-C_0 r^2} /A_{10} dr \notag \\
  & = \frac{1}{2 \sigma^2 (C_0 + \frac{T\sigma_n^2}{P_u G_0}) A_{10}}.
\end{align}
 $P^c_{C_{11}}$ and $ P^c_{C_{12}}$ can be obtained similarly.


\bibliographystyle{IEEEtran}
\bibliography{compresensive}

\end{document}